\RequirePackage{fix-cm}
\documentclass[smallextended]{svjour3}       
\smartqed  
\usepackage{graphicx}
\usepackage{tikz}
\usepackage{amsmath,amsfonts,amssymb,verbatim,float}
\usepackage{ mathrsfs }
\usepackage[utf8]{inputenc}
\usepackage{placeins}

\usepackage{float}
\restylefloat{table}

\usepackage{caption}
\usepackage{subcaption}
\newcommand{\mathbbm}[1]{\text{\usefont{U}{bbm}{m}{n}#1}}

\usepackage{overpic}

\begin{document}

\title{Time-varying volatility in Bitcoin market and information flow at minute-level frequency
\thanks{Authors would like to thank for the financial support from the EU Horizon 2020 SoBigData project under grant agreement No. 654024 and SoBigData++ with grant agreement 871042.}
}



\author{Irena Barjašić   \and
        Nino Antulov-Fantulin 
}


\institute{Irena Barjašić \at
              Department of Physics, Faculty of Science, University of Zagreb, 10000 Zagreb, Croatia \\         
           \and
           Nino Antulov-Fantulin \at
              ETH Zurich, Computational Social Science, 8092 Zürich, Switzerland\\
              \email{anino@ethz.ch}
}

\date{Received: date / Accepted: date}

\maketitle

\begin{abstract}
In this paper, we analyze the time-series of minute price returns on the Bitcoin market through the statistical models of the generalized autoregressive conditional heteroskedasticity (GARCH) family. 
We combine an approach that uses historical values of returns and their volatilities - GARCH family of models, with a so-called "Mixture of Distribution Hypothesis", which states that the dynamics of price returns are governed by the information flow about the market. Using time-series of Bitcoin-related tweets, the Bitcoin trade volume, and the Bitcoin bid-ask spread, as external information signals, we test for improvement in volatility prediction of several GARCH model variants on a minute level Bitcoin price time series. Statistical tests show that GARCH(1,1) and cGARCH(1,1) react the best to the addition of external signals to model the volatility process on out-of-sample data.\\
\keywords{Bitcoin \and Volatility \and Econometrics}
\end{abstract}

\section{Introduction}
\label{intro}

The first mathematical description of the evolution of price changes in a market dates back to Bachelier \cite{bachelier2011louis} (later rediscovered as Brownian motion, or random walk model), Mandelbrot \cite{Mandelbrot} (price increments are Lévy stable distribution), and truncated Lévy processes~\cite{stanley2000introduction}.
An opposing hypothesis (later named “Mixture of Distribution Hypothesis”) was introduced by Clark  \cite{Clark1973}, where the non-normality of price returns distribution is assigned to the varying rate of price series evolution during different time intervals. 
The process that is driving the rate of price evolution is proposed to be the information flow available to the traders. Due to the governing of the information flow, the number of summed price changes per observed time interval varies substantially, and the central limit theorem cannot be applied to obtain the distribution of price changes. Nevertheless, a generalization of the theorem provides a Gaussian limit distribution conditional on the random variable directing the number of changes~\cite{Clark1973}. 
In a  different approach, the autoregressive conditional heteroskedasticity (ARCH) \cite{ARCH} model, originally introduced by Engle, describes the heteroscedastic behavior (time-varying volatility) of logarithmic price returns relying only on the information of previous price movements. In addition to the previous values of price returns, its generalized variant GARCH \cite{GARCH} introduces previous conditional variances as well when calculating the present conditional variance.
GARCH is thus able to account for volatility clustering and for the leptokurtic distribution of price returns, both the stylized statistical properties of returns. 
An alternative view comes from GARCH-Jump model \cite{jorion1988jump}, which assumes that the
news process can be represented as $\epsilon_t=\epsilon_{1,t}+\epsilon_{2,t}$, a superposition of a normal component $\epsilon_{1,t}=\sigma_t z_t$ and a jump-like Poisson component with intensity $\lambda$. 
The constant intensity was generalized to autoregressive conditional jump intensity $\lambda_t=f(\lambda_{t-1})$ in \cite{chan2002conditional}. \\
Contrary to other studies about news jump dynamics and impact on daily returns \cite{chan2002conditional,maheu2004news}, we will model the volatility and external signals on a minute-level granularity. On this time-scale, our external signals are not modeled with Poisson-like dynamics, but added directly as an exogenous observable variable $I_{t-1}$ to form GARCHX model.\\
In this paper, we compare price volatility predictions of GARCH(1,1) with those of GARCHX(1,1) to explore how information is absorbed into the emerging cryptocurrency market of Bitcoin.  The Bitcoin \cite{BTCbook} is a cryptocurrency system operated through the peer-to-peer network nodes, with a public distributed ledger, called blockchain \cite{Nakamoto2008}. 
Similar to the foreign exchange markets, Bitcoin markets \cite{gandal2017price,ciaian2016economics} allow the exchange to fiat currencies and back.
Different studies on Bitcoin quantify the price formation \cite{cheah2015speculative,kristoufek2015main}, bubbles \cite{BouchaudBTC,sornetteBTC}, volatility \cite{Katsiampa2017,GuoBifetAntulov18}, systems dynamics \cite{Ron2013BTC,ElBahrawy2017,AntTol18} and economic value \cite{Hayes2015,Bolt2016,nadarajah2017inefficiency}. 
Although various studies \cite{kristoufek2013bitcoin,li2018sentiment,kim2016predicting,Garcia2015}  have used social signals from social media, WWW, search queries, sentiment, comments, and replies on forums, there still exists a gap in understanding Bitcoin volatility process through the autoregressive conditional heteroskedasticity models and exogenous signals. In this work we will focus on the statistical quantification 
the predictive power of the class of GARCH models with an external signal for the Bitcoin market on a minute level time-scale. 

\section{Data}
\begin{figure}[!htbp]
\begin{minipage}[b]{0.49\textwidth}
    \centering
    \includegraphics[width=1\textwidth]{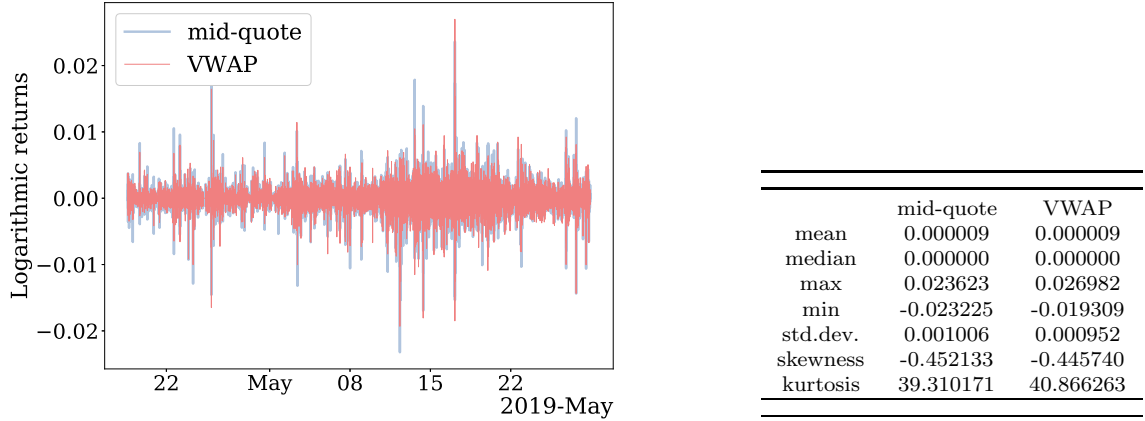}
    \par\vspace{0pt}
  \end{minipage}
  \hfill
  \begin{minipage}[b]{0.49\textwidth}
    \centering
\begin{tabular}{ccc}
\hline\noalign{\smallskip}
\noalign{\smallskip}\hline\noalign{\smallskip}
{} &  mid-quote &       VWAP \\
mean     &   0.000009 &   0.000009 \\
median   &   0.000000 &   0.000000 \\
max      &   0.023623 &   0.026982 \\
min      &  -0.023225 &  -0.019309 \\
std.dev. &   0.001006 &   0.000952 \\
skewness &  -0.452133 &  -0.445740 \\
kurtosis &  39.310171 &  40.866263 \\
\hline\noalign{\smallskip}
\noalign{\smallskip}\hline\noalign{\smallskip}
\end{tabular}
    \par\vspace{0pt}
    \end{minipage}
    \caption{ Volume weighted and mid quote logarithmic returns for the Bitcoin market. \textbf{a)} Time series. \textbf{b)} Descriptive statistics.}
    \label{fig:stat}
  \end{figure}

We used two types of price definitions, the mid-quote price and the volume weighted price, both calculated at a minute level.  Mid-quote price was constructed as the average between the maximum bid and minimum ask price on the last tick per minute, and the volume weighted average price (VWAP) as the volume weighted average of transaction prices per minute. \\
Sampling prices at such a high frequency brings up the issue of microstructure effects, such as bid-ask bounce, that introduces the autocorrelation between consecutive prices. Because of that, in addition to volume weighted prices, we use mid-quote prices that have a significantly smaller first order of autocorrelation,as explained in \cite{zero}, to strenghten the robustness of the results. An autocorrelation plot for both types of price returns is shown in the Appendix. \\
The Bitcoin prices were obtained from the Bitfinex exchange and logarithmic returns were calculated as a natural logarithm of two consecutive prices. The period we observed spans from April 18th, 2019 to May 30th, 2019, with 58 000 observations in total, 50 000 observations as in-sample and 8000 as out-of-sample, and is shown on Fig. \ref{fig:stat}a. In the table on Fig. \ref{fig:stat}b we can see the descriptive statistics of both kinds of logarithmic returns; the mean values of the returns are very close to zero ($8 \cdot 10^{-6}$), with standard deviations of $9.41\cdot 10^{-4}$ and $9.94\cdot 10^{-4}$, both distributions are negatively skewed and leptokurtic.


\begin{figure}[!htbp]
  \begin{minipage}[b]{0.49\textwidth}
    \centering
    \includegraphics[width=1\textwidth]{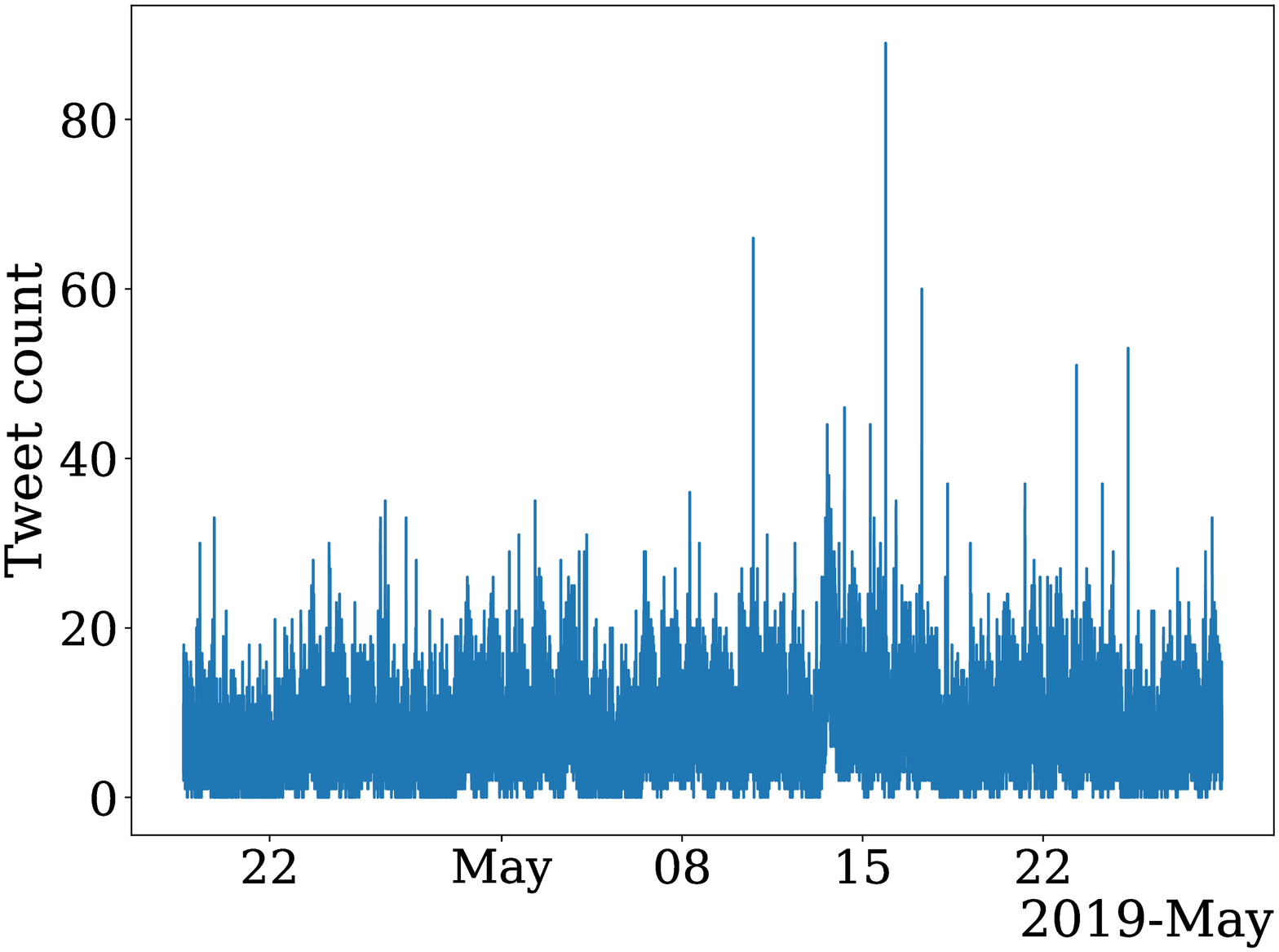}
        \par\vspace{0pt}
  \end{minipage}
  \hfill
  \begin{minipage}[b]{0.49\textwidth}
    \centering
    \includegraphics[width=1\textwidth]{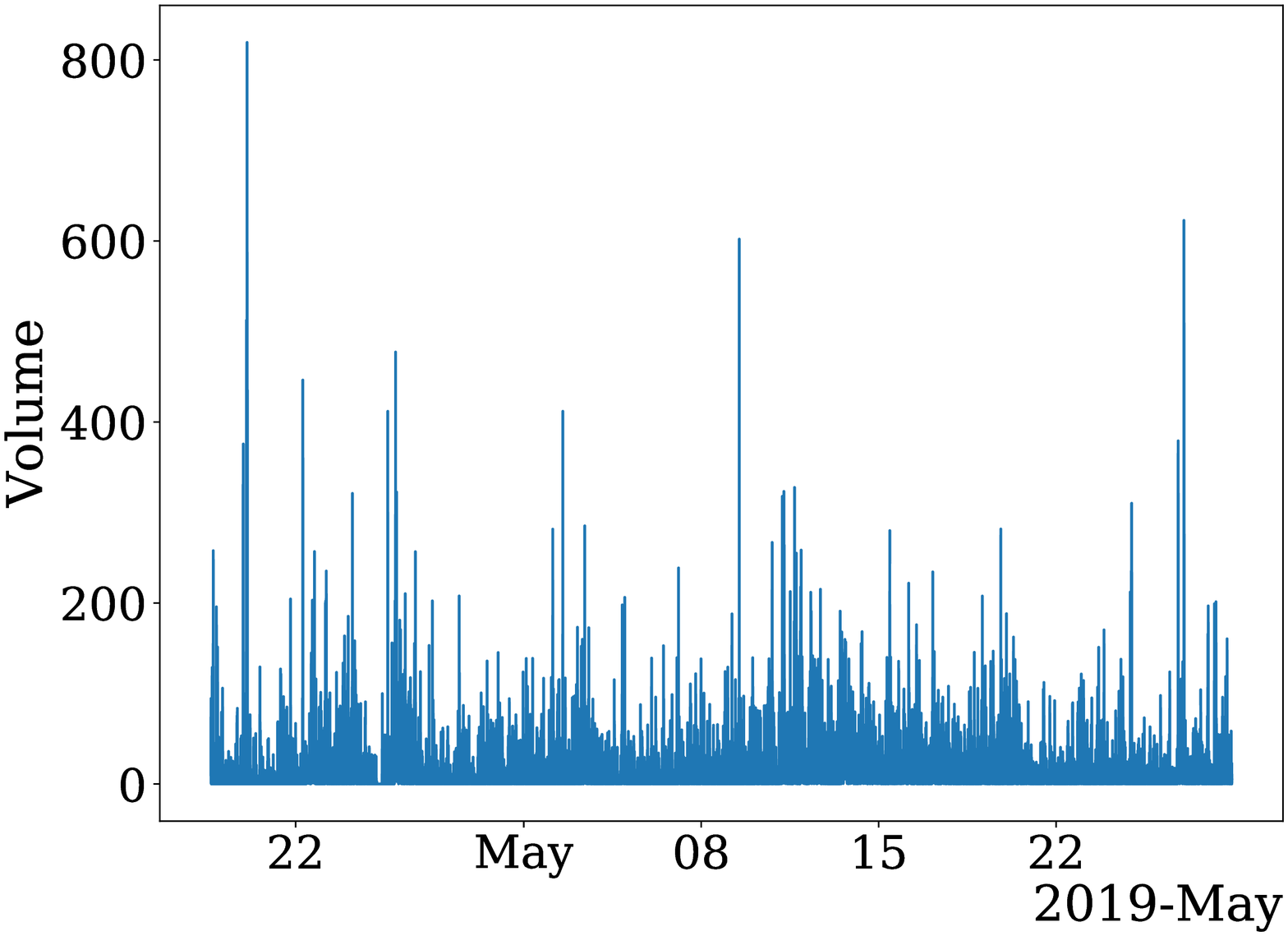}
        \par\vspace{0pt}
  \end{minipage}
  \begin{minipage}[b]{0.49\textwidth}
    \centering
    \includegraphics[width=1\textwidth]{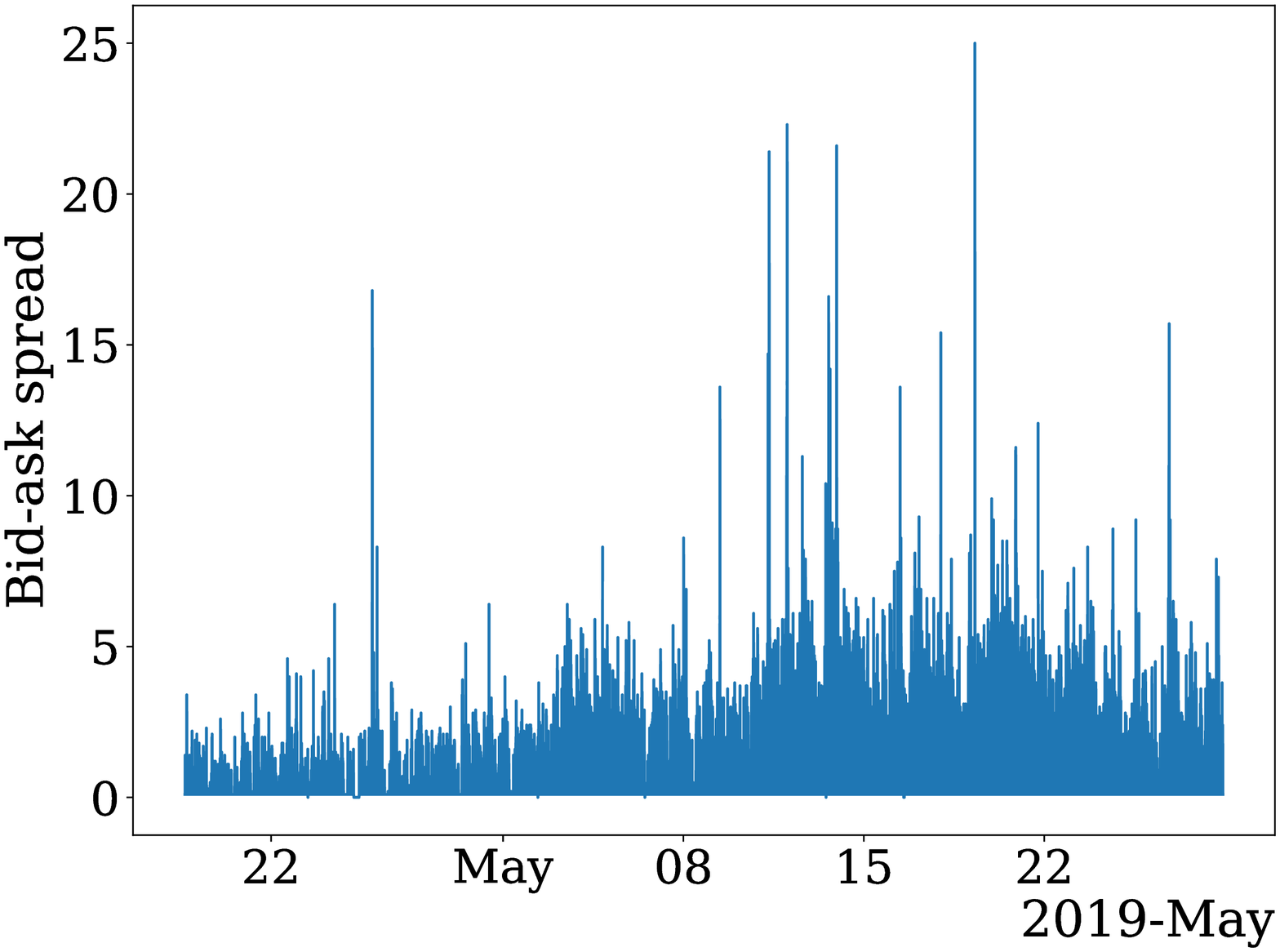}
        \par\vspace{0pt}
  \end{minipage}
  \hfill
  \begin{minipage}[b]{0.49\textwidth}
    \centering
\begin{tabular}{cccc}
\hline\noalign{\smallskip}
\noalign{\smallskip}\hline\noalign{\smallskip}
{} &  tweet count &  volume &  bid-ask \\
mean     &       6.7746 &    7.8801 &          0.4045 \\
median   &       6.0000 &    1.8624 &          0.1000 \\
max      &      89.0000 &  819.3810 &         25.0000 \\
min      &       0.0000 &    0.0000 &          0.0000 \\
std.dev. &       4.1544 &   18.4364 &          0.8846 \\
skewness &       1.5560 &    9.0434 &          5.2921 \\
kurtosis &       7.6529 &  186.7504 &         56.7640 \\
\hline\noalign{\smallskip}
\noalign{\smallskip}\hline\noalign{\smallskip}
\end{tabular}
    \par\vspace{0pt}
    \end{minipage}
    \caption{\textbf{a)} Time-series external signal of cryptocurrency-related tweets. \textbf{b)} Time-series of trading volume on Bitfinex market for BTC-USD pair. \textbf{c)} Time-series of bid-ask spread on Bitfinex market for BTC-USD pair. \textbf{d)} Descriptive statistics of external signals for Bitcoin market.}
    \label{fig:ext}
  \end{figure}

Three different datasets for external signals were available as the external information proxy - a time series of tweets mentioning cryptocurrency-related news \cite{beck2019sensing}, a time series of Bitcoin trade volumes from Bitfinex, and a time series of Bitcoin bid-ask spread, also from Bitfinex, all on a second level, shown on Fig. \ref{fig:ext}a,  Fig. \ref{fig:ext}b and Fig. \ref{fig:ext}c, with the descriptive statistics in Fig. \ref{fig:ext}d. All three time series were aggregated to the minute level.\\

\section{Mixture of distribution hypothesis}
\begin{figure}[!ht]
\centering
\begin{subfigure}{0.5\textwidth}
  \centering
  \includegraphics[width=1\textwidth]{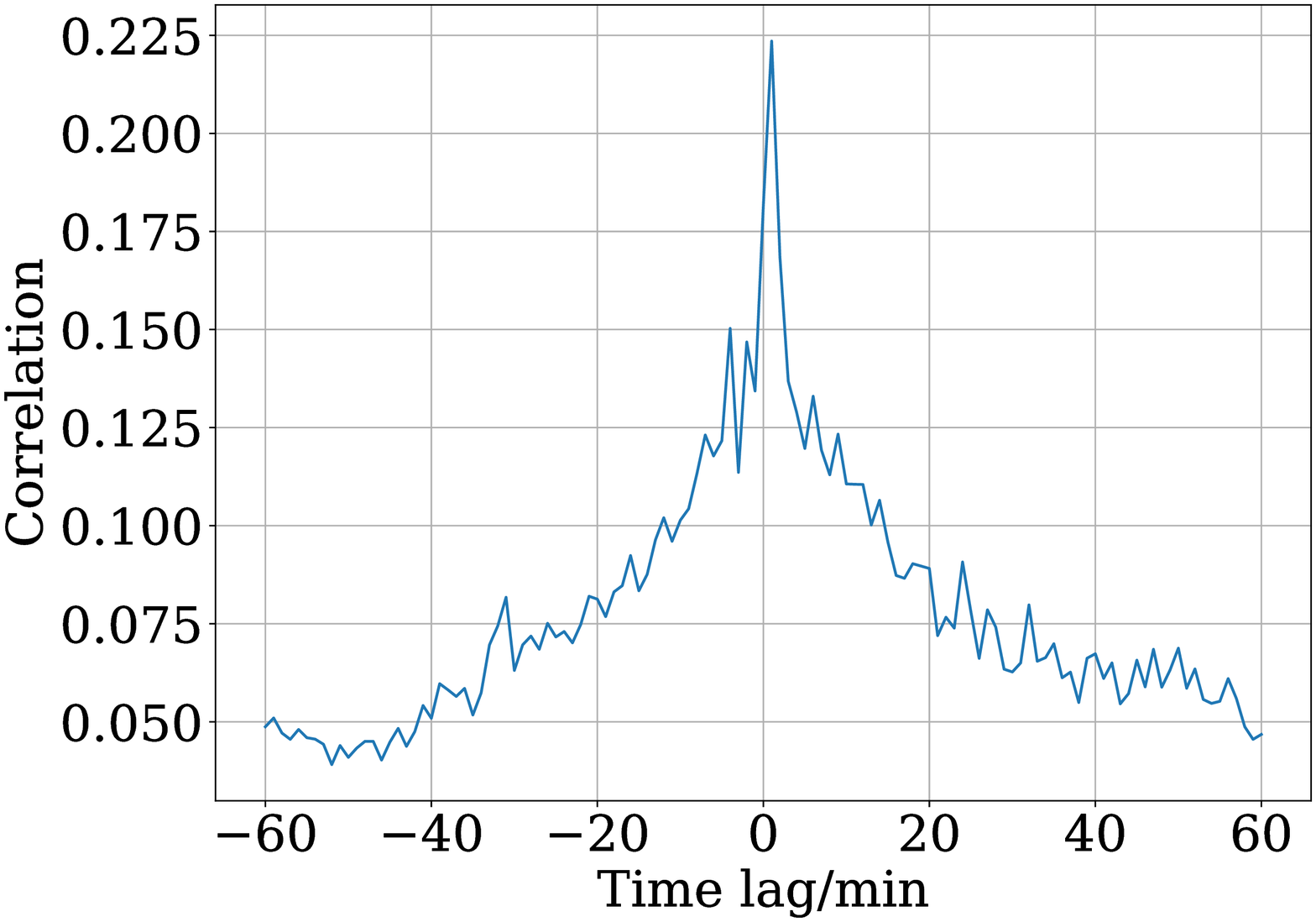}
  \caption{}
  \label{fig:vol_correlation}
\end{subfigure}%
\begin{subfigure}{0.5\textwidth}
  \centering
  \includegraphics[width=1\textwidth]{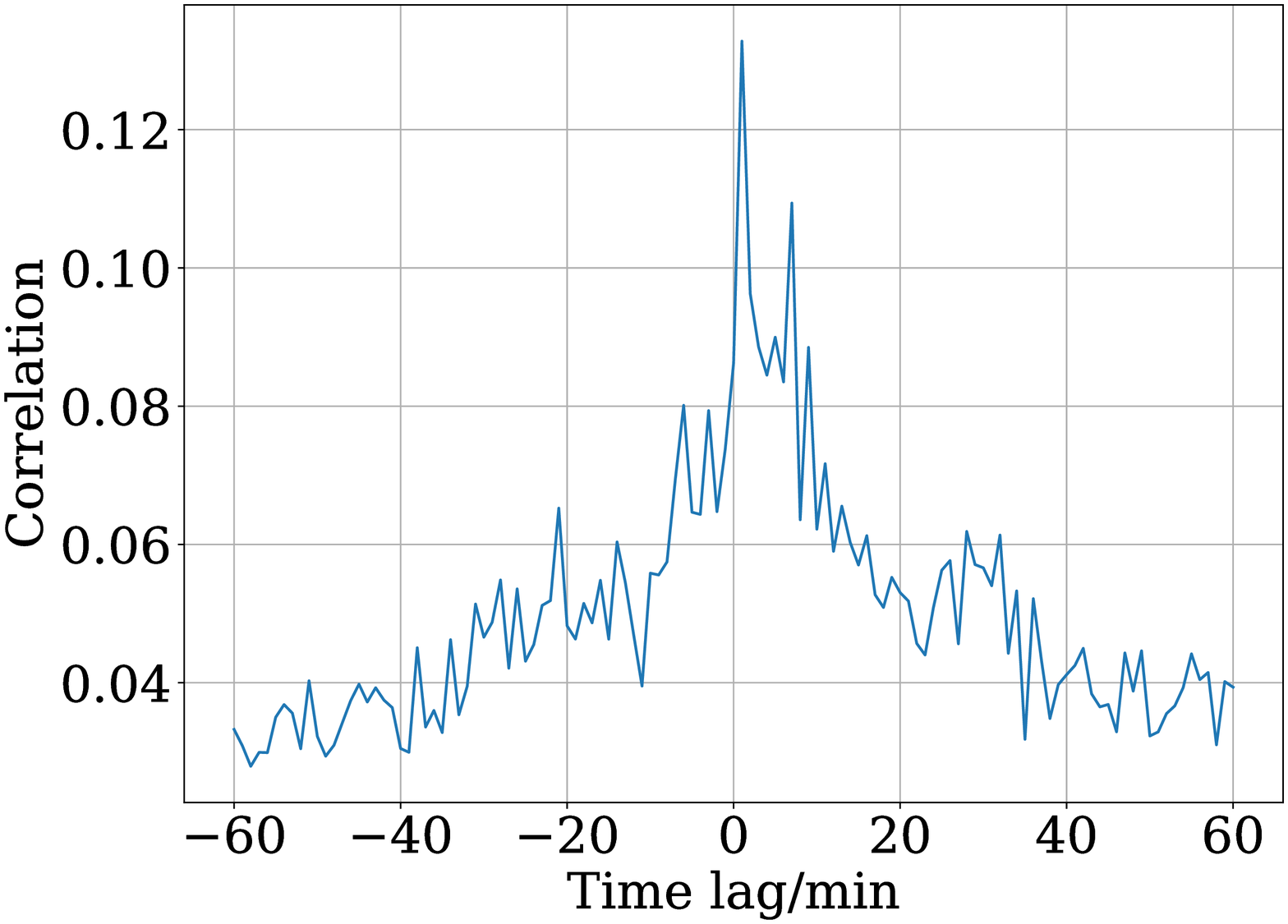}
  \caption{}
  \label{fig:BA_correlation}
\end{subfigure}
\caption{(a) Squared volume-weighted price returns - volume correlation. All values of correlation are statistically significant (p-value $ \leq 0.001$). Permutation significance check indicates no statistically significant correlation between time-permuted squared price returns and volume series. (b) Squared volume-weighted price returns - bid-ask spread correlation. All values of correlation are statistically significant (p-value $ \leq 0.001$). Permutation significance check indicates no statistically significant correlation between time-permuted squared price returns and volume series.).
}
\end{figure}

The “Mixture of Distribution Hypothesis” models the non-normality of price returns distribution with the varying rate of price series evolution due to the different information flow during different time intervals.
Practically, Clark \cite{Clark1973} hypothesizes that this can be observed as a linear relationship between the proxy for the information flow $I_t$ and the price change variance $r_t^2$, and suggests trading volume $v_t$ as the proxy. Tauchen and Pitts \cite{Tauchen} state a bivariate normal mixture model which conditions the price returns and trading volume on the information flow as: 
\begin{align}
  	r_t &= \sum_{i=1}^{I_t} r_{t,i}, \quad r_{t,i} \in \mathcal{N}(0,\sigma_1).\\
    v_t &= \sum_{i=1}^{I_t} v_{t,i}, \quad v_{t,i} \in \mathcal{N}(\mu_2,\sigma_2).
\end{align}
Both, the price return and trading volume are mixture of independent normal distributions with the same mixing variable  $I_t$, which represents the number of new pieces of information arriving to market.
Conditioned on $I_t$, price changes are distributed as $\mathcal{N}(0,I_t\sigma_1)$ and trading volume is distributed as $\mathcal{N}(I_t \mu_2,I_t \sigma_2)$, and the model can be rewritten as:
\begin{align}
\label{eq:mdh}
  	r_t &= {\sigma}_1 \sqrt{I_t} z_{1t}, \quad z_{1t} \in N(0,1).\\
    v_t &= {\mu}_2 I_t + {\sigma}_2 \sqrt{I_t} z_{2t}, \quad z_{2t} \in N(0,1).
\end{align}
The relationship between price variance and trading volume immediately follows:
\begin{equation}
    Cov({r_t}^2,v_t) = {\sigma}_1 {\mu}_2 Var(I_t),
\end{equation}
and the stochastic term in (4) shows that the above proposed linear relationship is only an approximation.\\
To start our analysis, we calculated correlation plots for the relationship between the external signals and the squared VWAP price returns.
Correlation between squared price returns and volume was calculated for different time lags of the volume time series, as shown in Fig. \ref{fig:vol_correlation}. Both have a peak when the external series leads the squared price returns by one minute. The significant correlation, i.e. normalized covariance between squared price returns and trading  volume, indicates an approximately linear relationship between the volatility and the two proxies for information flow (see Eq. 3). The result we got using the bid-ask spread as external signal can be seen (Fig. \ref{fig:BA_correlation}) to be analogous to the one obtained for volume.\\

\begin{figure}[!ht]
\centering
\begin{subfigure}{0.5\textwidth}
  \centering
  \includegraphics[width=1\linewidth]{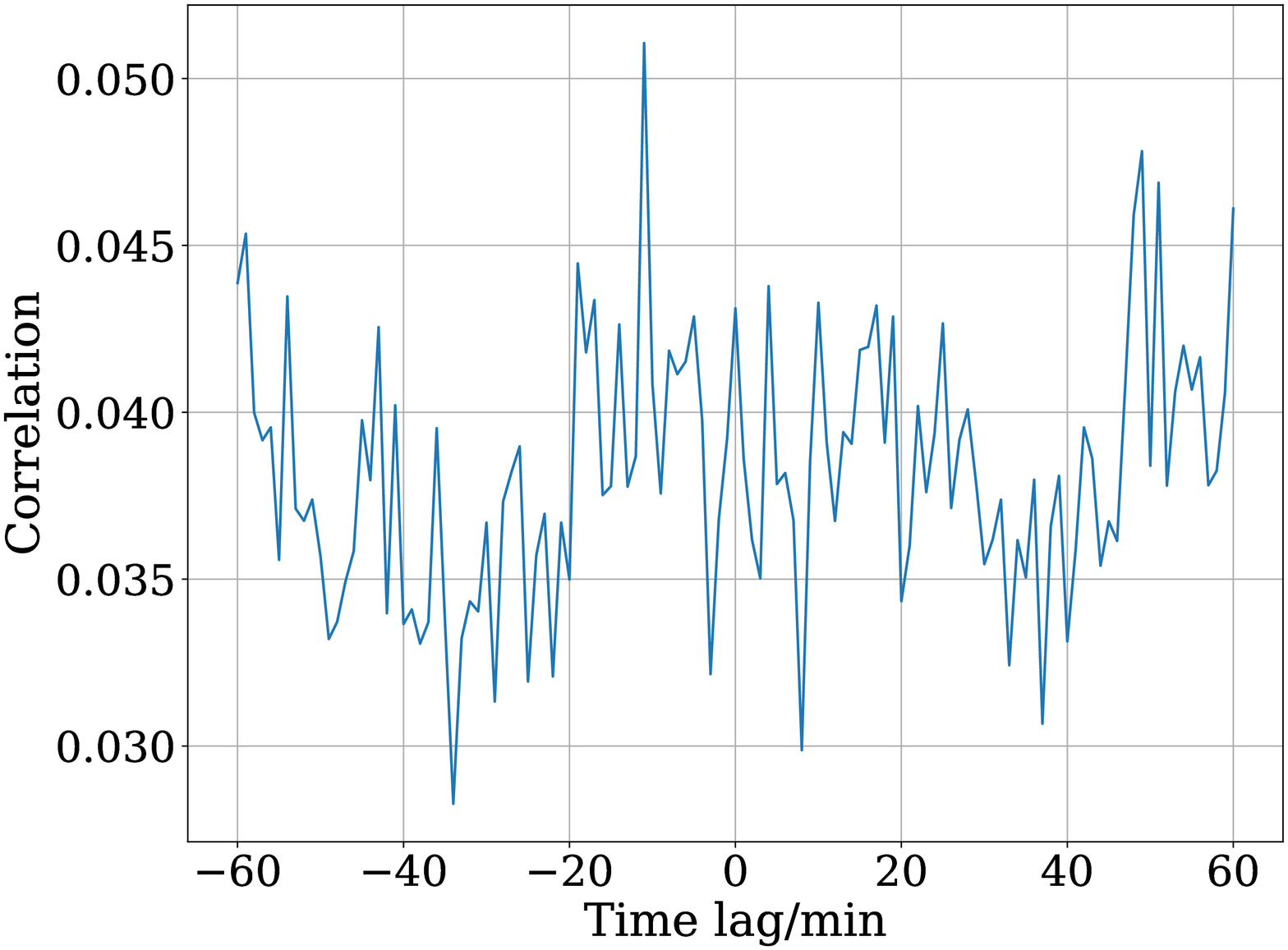}
  \caption{}
  \label{fig:tweet_correlation}
\end{subfigure}%
\begin{subfigure}{0.5\textwidth}
  \centering
  \includegraphics[width=1\linewidth]{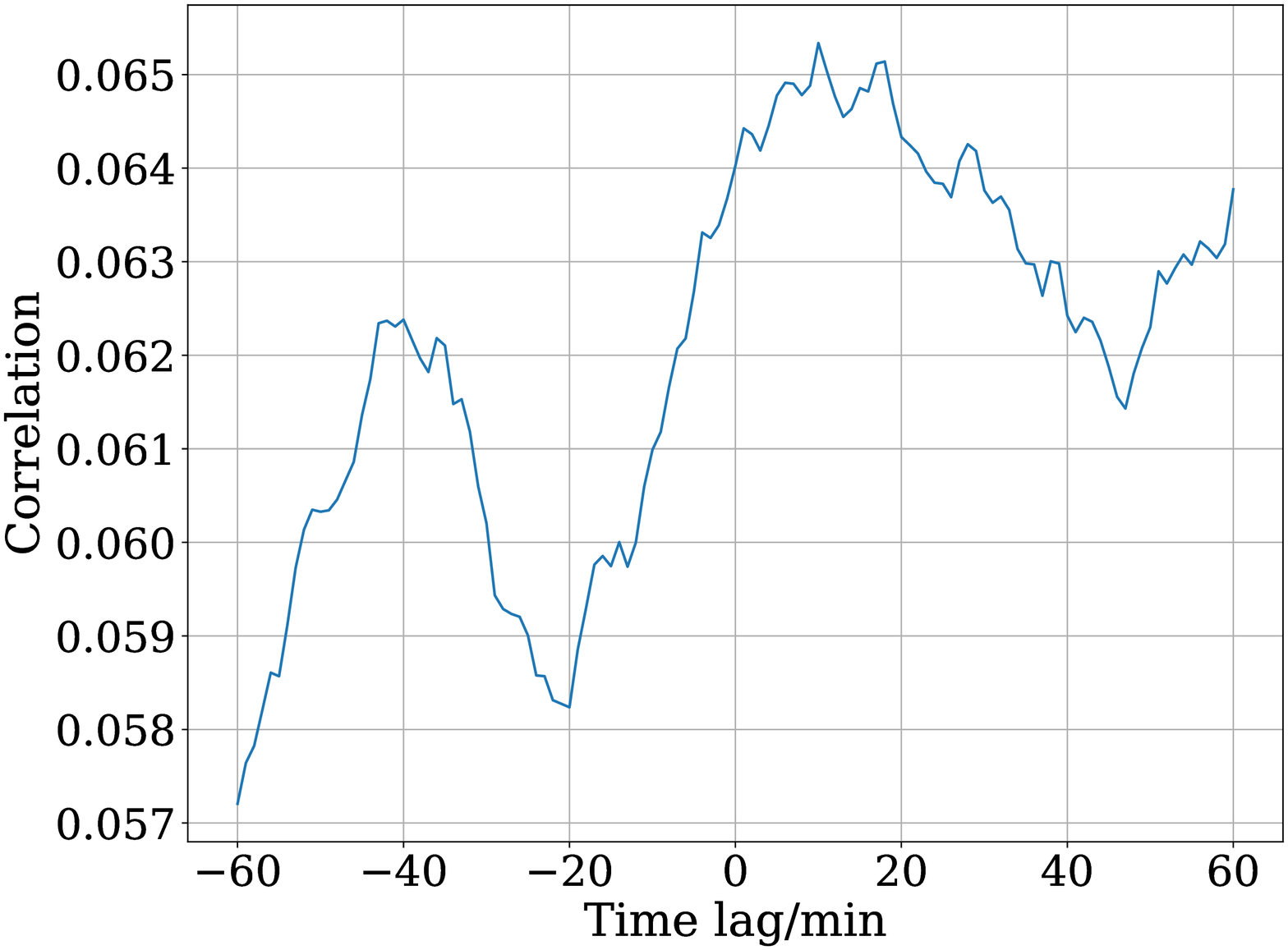}
  \caption{}
  \label{fig:tweet_int_correlation}
\end{subfigure}
\caption{(a) Correlation between squared price returns and Twitter volume. Permutation significance check indicates no statistically significant correlation between time-permuted squared price returns and Twitter time series. (b) Correlation between squared price returns and integrated Twitter volume (over a 30-minute moving window). This test is only used to check whether the integrating operator is filtering noise. 
Correlation between squared price returns and Twitter time series. All values of correlation are statistically significant (p-value $ \leq 0.001$).
}
\end{figure}

In Appendix we plot the same correlation calculation for cryptocurrency-related tweets (see Fig. \ref{fig:tweet_correlation}). We do not observe a similar correlation (covariance) pattern as for volume and bid-ask spread signals. 
Multiple reasons could be behind this: (i) a large noise in the Twitter signal might be covering the information flow w.r.t. trading volume signal, (ii) linear dependence might not be enough to capture the relationship, or (iii) Twitter signal might not contain a sufficient information flow to influence price volatility. If noise is i.i.d., then "integrated external signal" $\Tilde{I}(t)=\int_{t-\delta}^{t} I_t dt$ should filter the noise component. 
 We observe that the stronger correlation pattern is present after the Twitter series is integrated with $\delta=30$ minutes (see Appendix Fig. \ref{fig:tweet_int_correlation}), which indicates that strong noise is present in Twitter series.
 
\section{Transfer entropy between information flow and volatility proxy}
\begin{figure}
\centering
  \includegraphics[width=0.8\textwidth]{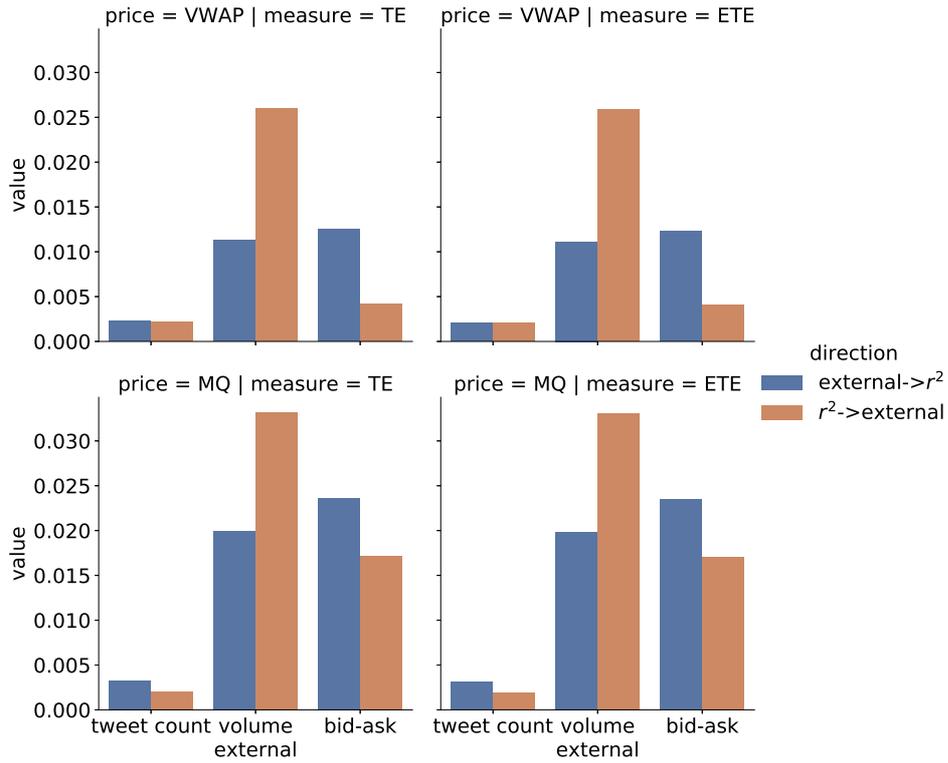}
\caption{Transfer entropy (TE) and effective transfer entropy (ETE) between external signals (Twitter, volume and bid-ask spread) and squared returns (VWAP and mid-quote price returns). All transfer entropy results are statistically significant (p-value smaller than 0.001), additionally the presence of unit-roots was checked with augmented Dickey–Fuller test ($\alpha=0.01$). }
\label{fig:TE}       
\end{figure}

To proceed, we move from the linear dependence that is captured with correlation $\rho({r_t}^2,v_t)$ to check the non-linear dependence argument between the squared returns and external information flow $I_t$ signals (volume, bid-ask spread and Twitter) in causal setting ${r_t}^2=f(I_{t-1}, r_{t-1})$. 
In particular, for the squared price return process $\{ r_t^2 \}$ and external information proxy process $\{ I_t \}$, we calculate \textbf{transfer entropy} (TE) \cite{Schreiber_2000}.
\begin{equation}\label{eq:transfer-entropy}
  TE_{I \to r^2} := H(r^2_{t+1}| r^2_t ) - H(r^2_{t+1}| r^2_t, I_t), 
\end{equation}
where $H(X|Y):=-\sum_{i,j} p(x_i,y_j) \log[p(x_i|y_j)]$ 
denotes the conditional Shannon entropy.
Transfer entropy is an information-theoretic measure that is both nonlinear and non-symmetric, and it does not require a Gaussian assumption for the time series \cite{Barnett_2009}. The non-symmetry allows us to distinguish the direction of information exchange between time series, $I_t$ and $r^2_t$. 
In Fig. \ref{fig:TE} we present the results for transfer entropy from external variables to squared returns time series and conversely. The stationarity of the series was checked using ADF test and the hypothesis of the unit root was rejected at a $1 \%$ significance. Results of the transfer entropy analysis show that values are significant, with the largest one being the transfer entropy from squared returns to trading volume. The statistical significance (p-value) of transfer entropy was estimated by a bootstrap method of the underlying Markov process \cite{TEpval}. To account for the finite sample size, we use the effective transfer entropy (ETE) measure:
\begin{equation}
ETE_{I \to r^2} =  T_{I \to r^2} - \frac{1}{M} \sum_{m=1}^M T_{I_{(m)} \to r^2},
\end{equation}
where $I_{(m)}$ is the $m$-th shuffled series of $I$ \cite{ETE}.
We observe stronger information transfer from the volume signal and the bid-ask spread to squared returns than from the Twitter signal to squared returns. At this point, we conclude that all external signals show significant dependence towards the proxy for volatility signal i.e. squared returns. 


\section{Generalized autoregressive conditional heteroskedasticity with external information flow}

Using the transfer entropy analysis we have found statistically significant dependence between historical information proxy and volatility proxy, but not the actual functional dependence. 
Therefore, we now turn to the class of generalized autoregressive conditional heteroskedasticity models~\cite{GARCH}, that well describe the price return process and augment it with the external information flow proxy signal. 

The GARCH(1,1) model  conditions the volatility on its previous value and the previous value of price returns:
\begin{align}
  	r_t & = {\mu}_t + {\varepsilon}_t, \quad  {\varepsilon}_t = {\sigma}_t z_t, \quad z_t \in N(0,1).\\
    {\sigma}^2_t &= \omega + \alpha {\varepsilon}^2_{t-1} + \beta {\sigma}^2_{t-1}.
\end{align}
Large $\alpha$ coefficient indicates that the volatility reacts intensely to market movements, while large $\beta$ shows that the impact of large volatilities slowly dies out. The volatilities defined by the model display volatility clustering and the respective distribution of price returns is leptokurtic, which agrees with the observations in the real data.\\


Motivated by MDH and TE analysis, we formed a GARCHX model by adding the proxy for the information flow  $I_{t-1}$ directly to the GARCH volatility equation
\begin{equation}
    {\sigma}^2_t = \omega + \alpha {\varepsilon}^2_{t-1} + \beta {\sigma}^2_{t-1} + \gamma I_{t-1}.
\end{equation}
We will compare price volatility predictions of GARCH(1,1) with those of GARCHX(1,1) to explore how information is absorbed into the emerging cryptocurrency market of Bitcoin.



\subsection{Volatility GARCHX process analysis}
\begin{figure}
\centering
  \includegraphics[width=0.8\textwidth]{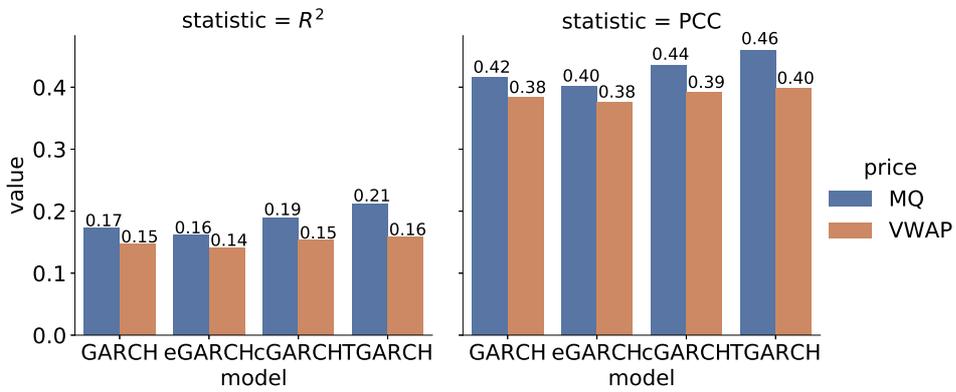}
\caption{Out-of-sample measures for the GARCH volatility process. In-sample consists of 50 000 points, out-of-sample consists of 8000 points.
All PCC values are statistically significant. R$^2$ statistic significance was checked using F-statistic, and satisfied for all the values.}
\label{fig:R_PCC}       
\end{figure}

We turn our attention to the statistical quantification of the GARCH volatility processes. Apart from expanding GARCH(1,1) to GARCHX(1,1), we add the exogenous variable to models eGARCH(1,1), cGARCH(1,1)  and TGARCH(1,1) as well, to check for improvement in volatility predictions. The conditional variance equations corresponding to these models (see Table \ref{tab:GARCH}) are extensions of Eq. 5. eGARCH \cite{nelson1991eGARCH} and TGARCH \cite{TARCH} capture the asymmetry between positive and negative shocks, giving greater weight to the later ones, with the difference between them being the multiplicative and the additive contribution of historical values, and cGARCH \cite{1999cGARCH} separates long and short-run volatility components.
\begin{table}[ht]
\centering
\caption{GARCH family}
\label{tab:GARCH}       
\begin{tabular}{c} 
\hline
\hline \\
eGARCH \\
$\ln({\sigma}_t^2) = \omega + \alpha \left[ \left| \frac{{\varepsilon}_{t-1}}{{\sigma}_{t-1}} \right| - E\left| \frac{{\varepsilon}_{t-1}}{{\sigma}_{t-1}} \right|  \right] + \delta \frac{{\varepsilon}_{t-1}}{{\sigma}_{t-1}} + \beta \ln ({\sigma}_{t-1}^2)$\\ \\
\hline \\
cGARCH \\
${\sigma}_t^2 = q_t + \alpha ({\varepsilon}^2_{t-1} - q_{t-1})+ \beta ({\sigma}_{t-1}^2 - q_{t-1})$ \\
        $q_t = \omega  + \rho q_{t-1} + \theta ({\varepsilon}^2_{t-1} - {\sigma}^2_{t-1})$\\ \\
\hline \\
TGARCH \\
${\sigma}_t = \omega + \alpha {\varepsilon}_{t-1} + \beta {\sigma}_{t-1} + \phi {\varepsilon}_{t-1} \mathbbm{1}_{[\varepsilon_{t-1}<0]}$ \\ \\
\hline
\hline
\end{tabular}
\end{table} \\
To get the intuition how good the GARCH volatility models are at explaining the volatility, we regress $a \cdot \sigma_t^2+b$ on squared returns $r_t^2$ \cite{andersen1998answering}, where $\sigma_t^2$ is the squared GARCH volatility estimate (out-of-sample). Then, we measure the coefficient of determination $R^2$ i.e. the proportion of the variance in the dependent variable that is predictable from the independent variable. We determine the statistical significance of $R^2$ with the F-test. Additionally, we measure the Pearson correlation coefficient (PCC) of estimated ${\sigma}_t^2$ and squared returns $r_t^2$, along with its statistical significance, Fig. \ref{fig:R_PCC}. 

\begin{figure}
\centering
  \includegraphics[width=0.8\textwidth]{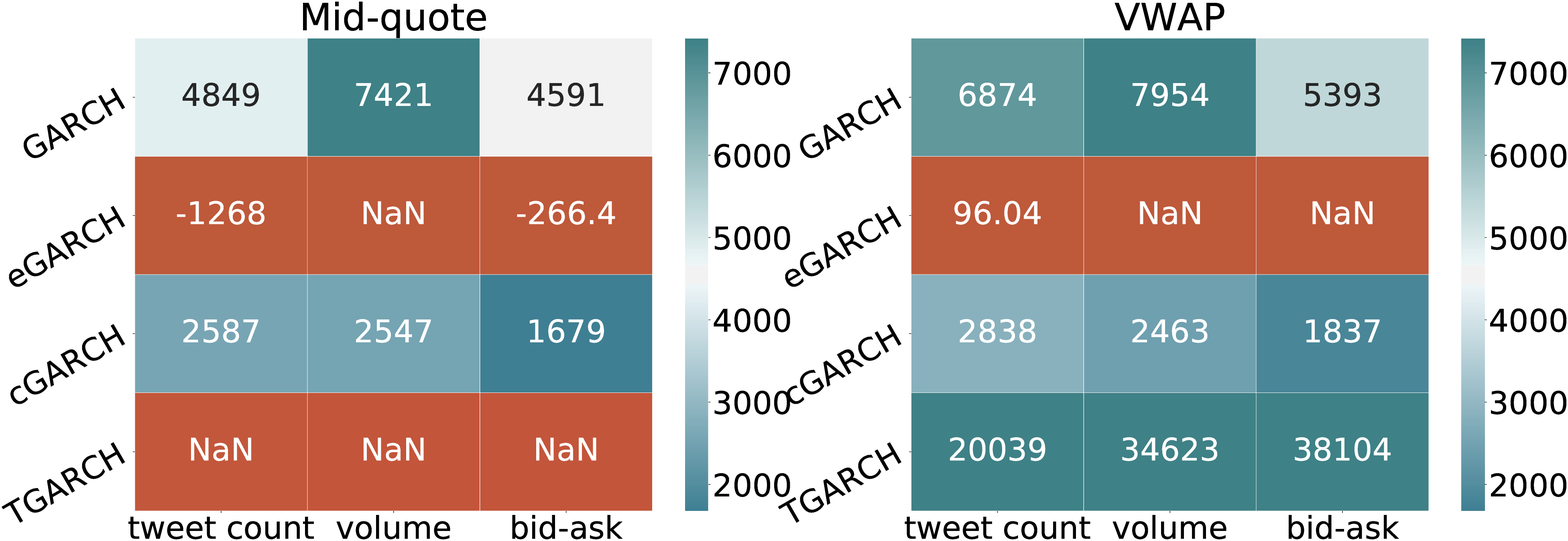}
\caption{Results of out-of-sample likelihood ratio test. In-sample consists of 50 000 points, out-of-sample consists of 8000 points.\\ $^{*}$Blue palette represents the p-value smaller than 0.001. NaN - some algorithms had convergence problems.}
\label{fig:LLR}       
\end{figure}

However, for a more precise statistical quantification of the difference between models and their GARCHX variants more advanced statistical tests are needed. For that purpose, we employ predictive negative log-likelihood (NLLH) \cite{wu2014gaussian}. 
\begin{equation}
    \Tilde{\mathcal{L}} =-\ln (\mathcal{L}({\mu}_1,...,{\mu}_n,{\sigma}_1,...,{\sigma}_n)) = -\sum_{i = 1}^n \left( \frac{1}{2} \ln({\sigma}_i) + \frac{1}{2}\ln(2 \pi) - \frac{{(r_i-{\mu}_i})^2}{2 {\sigma}_i^2} \right).
\end{equation}
We evaluated predictive negative log-likelihood (NLLH) on the out-of-sample period. Values of $\{{\mu}_i\}_{i=1}^n$ and $\{{\sigma}_i\}_{i=1}^n$ are predictions of the model, and $\{r_i\}_{i=1}^n$ are observed price returns. To show whether the improvements can be considered significant, we employed the likelihood ratio test. It takes the natural logarithm of the ratio of two log-likelihoods as the statistic: 
\begin{equation}
    LR = -2 \ln \left( \frac{\mathcal{L}({\theta}_0)}{\mathcal{L}(\theta)} \right).
\end{equation}
Since its asymptotic distribution is ${\chi}^2$-distribution, a p-value is obtained using Pearson's chi-squared test.
In Fig. \ref{fig:LLR}, we see from the p-values that the exogenous variables improve the NLLH significantly for all the models except for eGARCH, for logarithmic returns are created from VWAP. When mid-quote prices are used, significant improvement is observed only for GARCH and cGARCH.\\
Note, that for two models with fixed parameters, the likelihood ratio test is the most powerful test at given significance level $\alpha$, by Neyman–Pearson lemma. \\
In order to further test the \textbf{robustness} of the conclusions on different samples, we perform the \textbf{bootstrapping}. We restrict the lengths of in-sample and out-of-sample to $T=1000$ points each and bootstrap $N=100$ such segments from the original time series. Then, for each segment we fit a model on its in-sample data segment and calculate predictive out-of-sample NLLH $\{\Tilde{\mathcal{L}}_i\}_{i=1}^N$. 
\begin{figure}
\begin{subfigure}{.5\textwidth}
  \centering
  \includegraphics[width=1\linewidth]{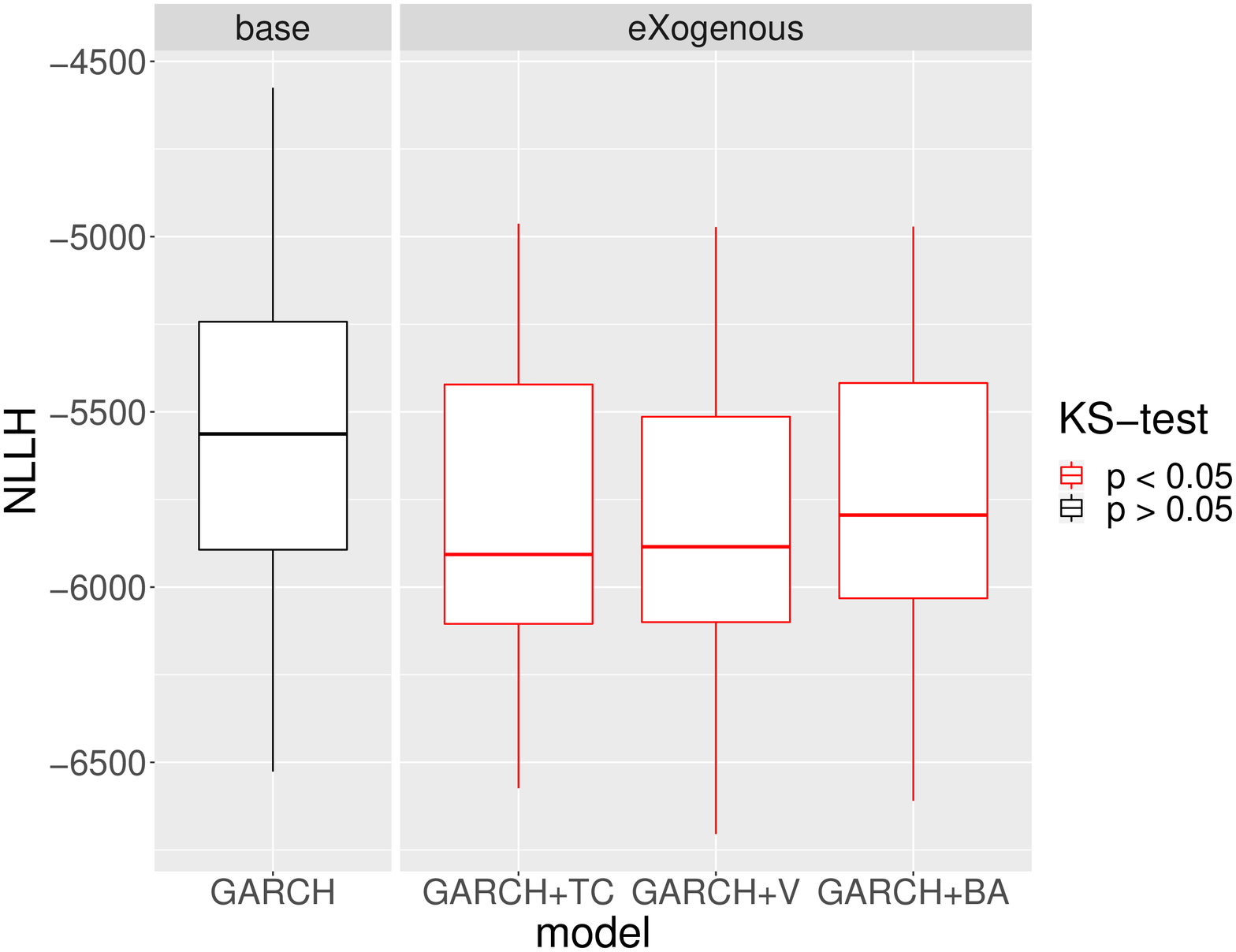}  
\end{subfigure}
\begin{subfigure}{.5\textwidth}
  \centering
  \includegraphics[width=1\linewidth]{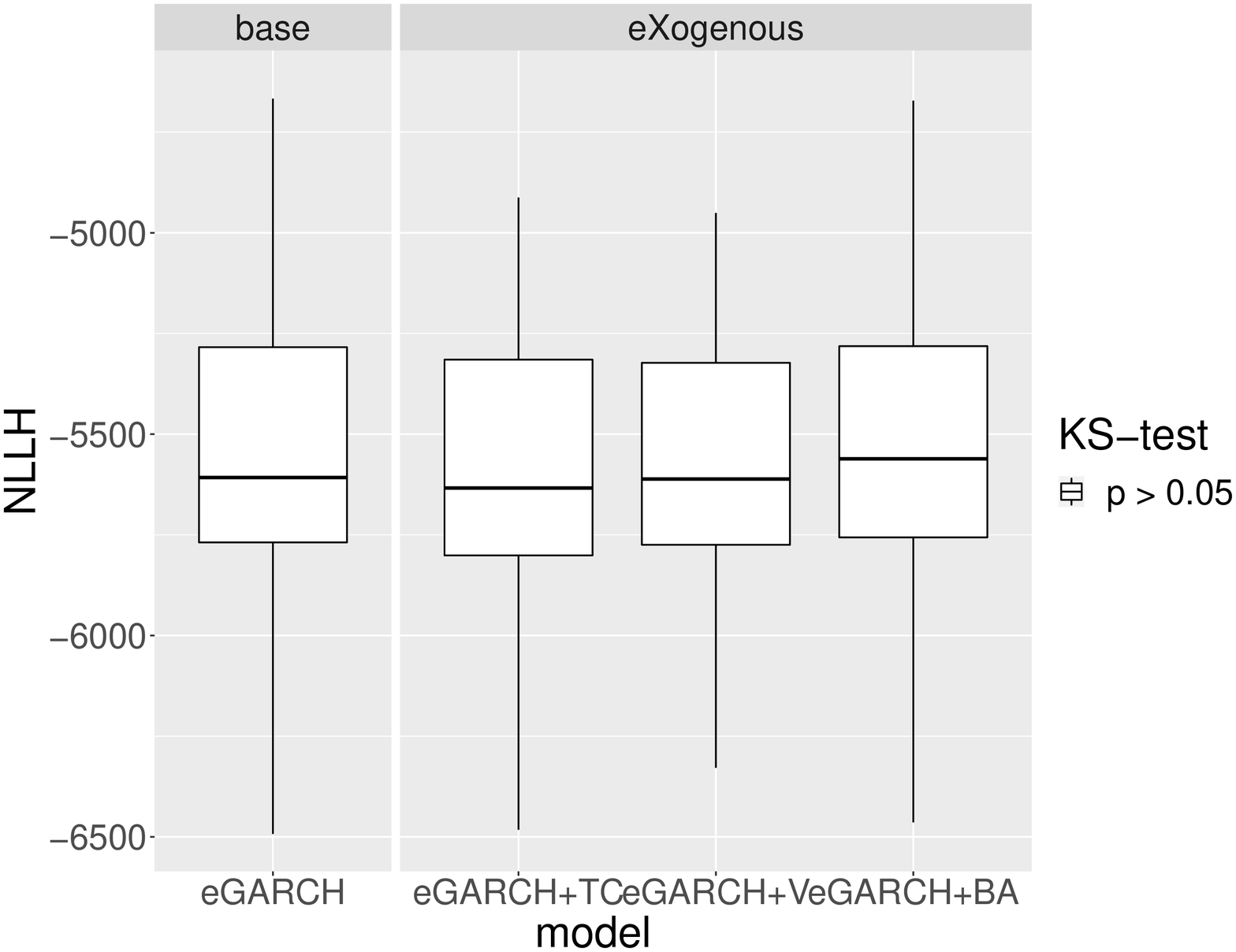}  
\end{subfigure}
\begin{subfigure}{.5\textwidth}
  \centering
  \includegraphics[width=1\linewidth]{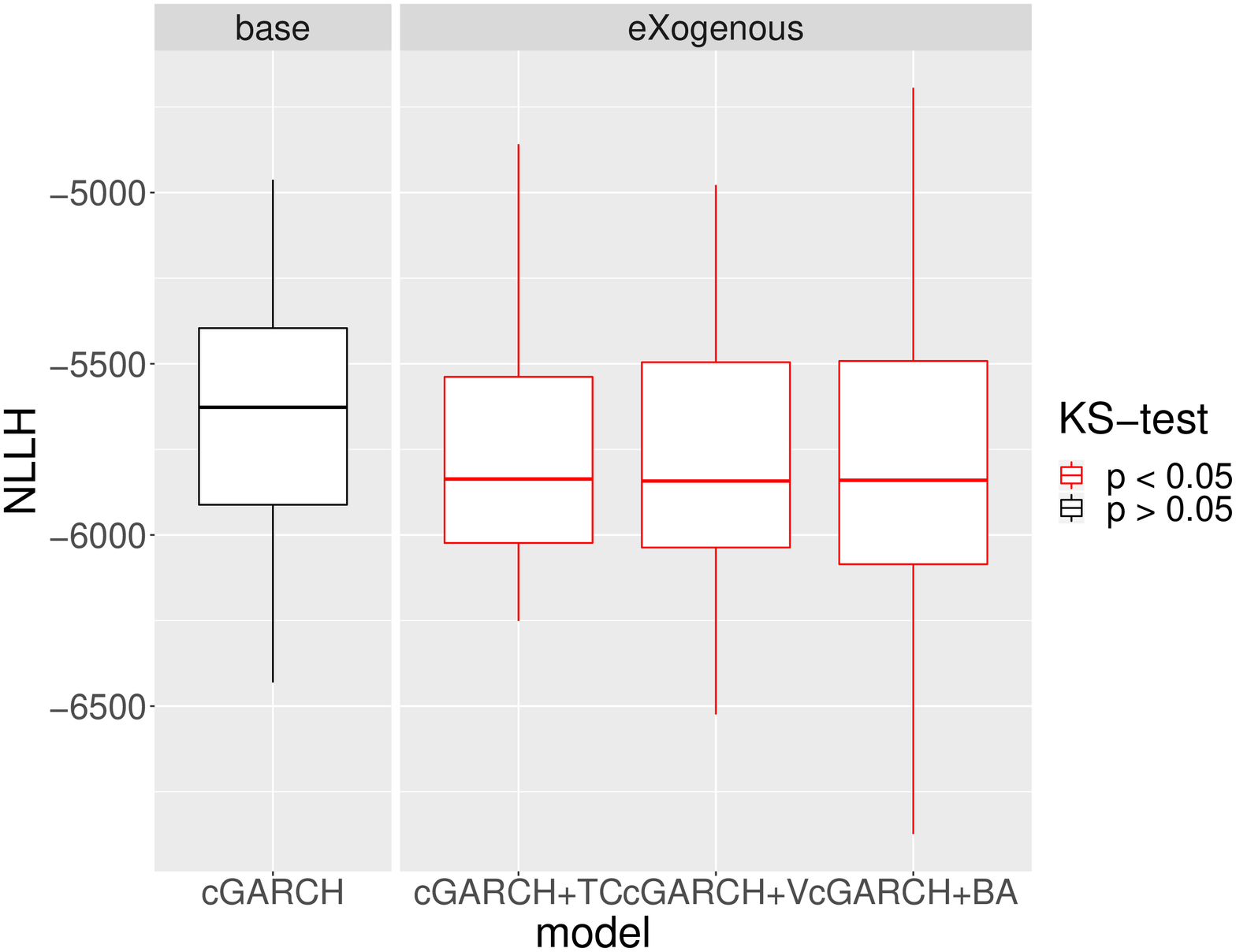}  
\end{subfigure}
\begin{subfigure}{.5\textwidth}
  \centering
  \includegraphics[width=1\linewidth]{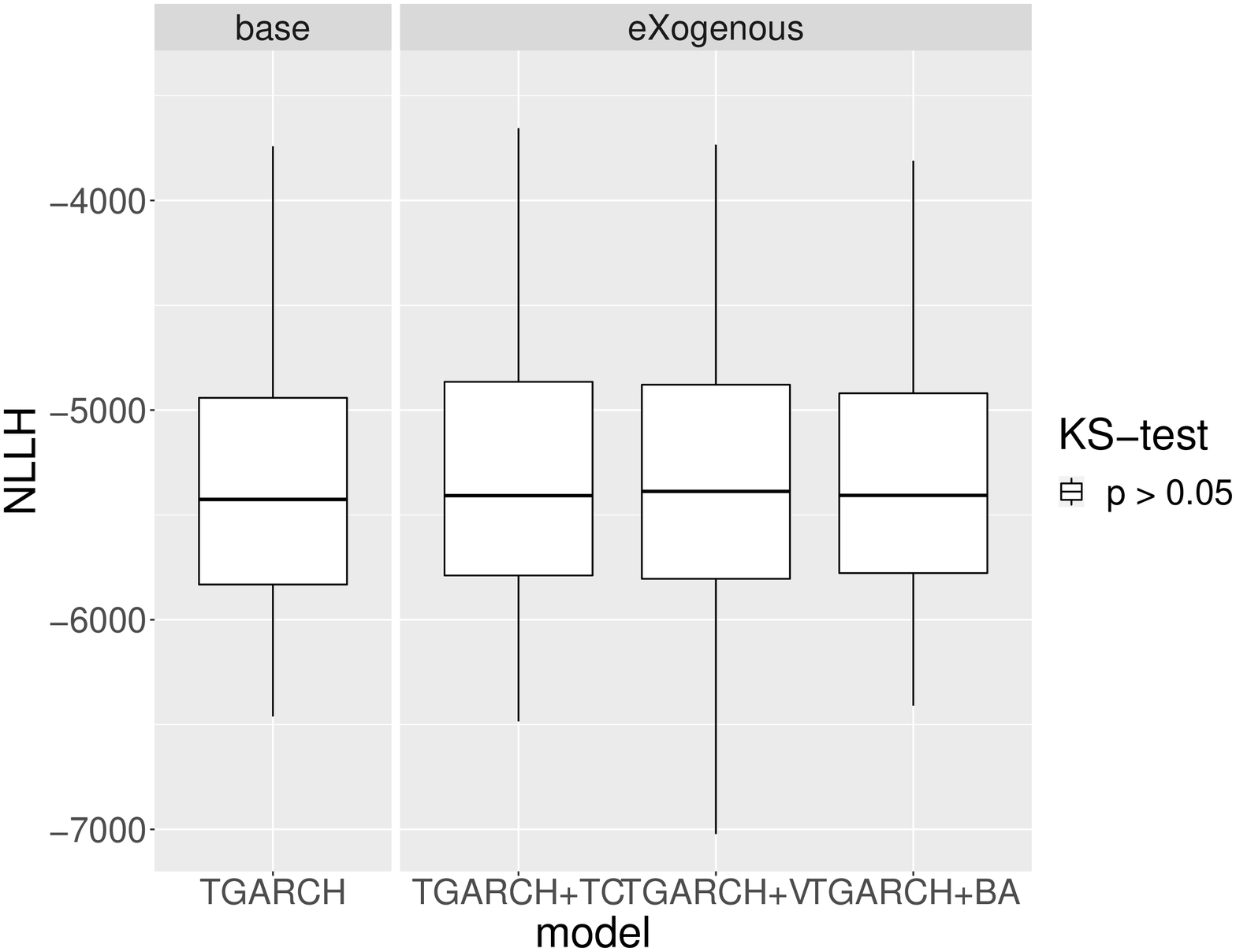}  
\end{subfigure}
\caption{Bootstrap robustness check over N = 100 splitting points with T = 1000 training points and T = 1000 test size for GARCH and GARCHX models. The price is defined as volume-weighted. The non-parametric Kolmogorov–Smirnov test of the equality of the NLLH out-of-sample distributions between the GARCH and GARCHX models is done. \textbf{a)} KS test implies a significant difference for both external signals for the GARCH model. \textbf{b)} KS test implies no significant difference for external signals for eGARCH model. \textbf{c)} KS test implies no significant difference for both external signals for the cGARCH model. \textbf{d)} KS test implies no significant difference for external signals for the TGARCH model.}
\label{fig:vwapnllh}
\end{figure}
\begin{figure}
\begin{subfigure}{.5\textwidth}
  \centering
  \includegraphics[width=1\linewidth]{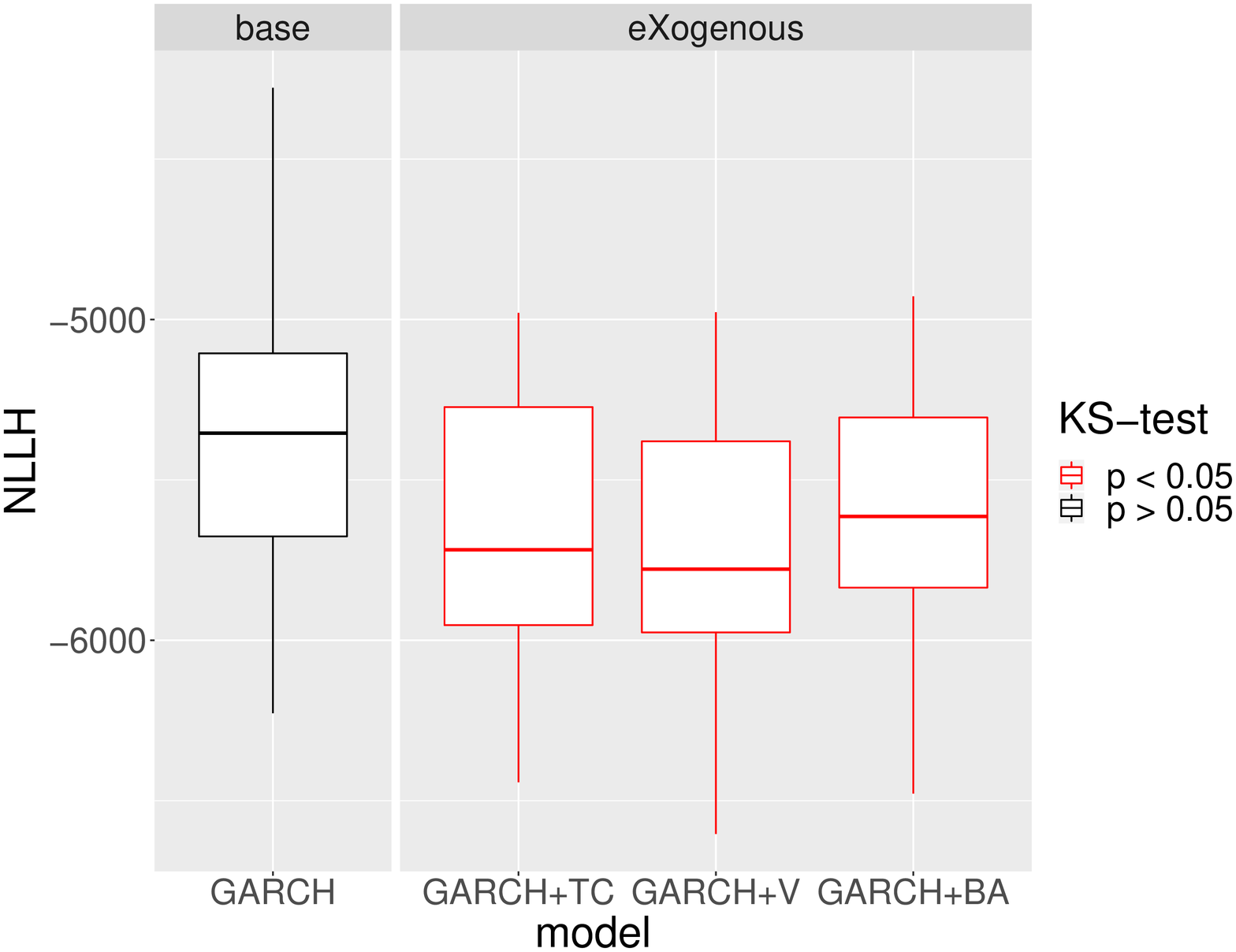}  
\end{subfigure}
\begin{subfigure}{.5\textwidth}
  \centering
  \includegraphics[width=1\linewidth]{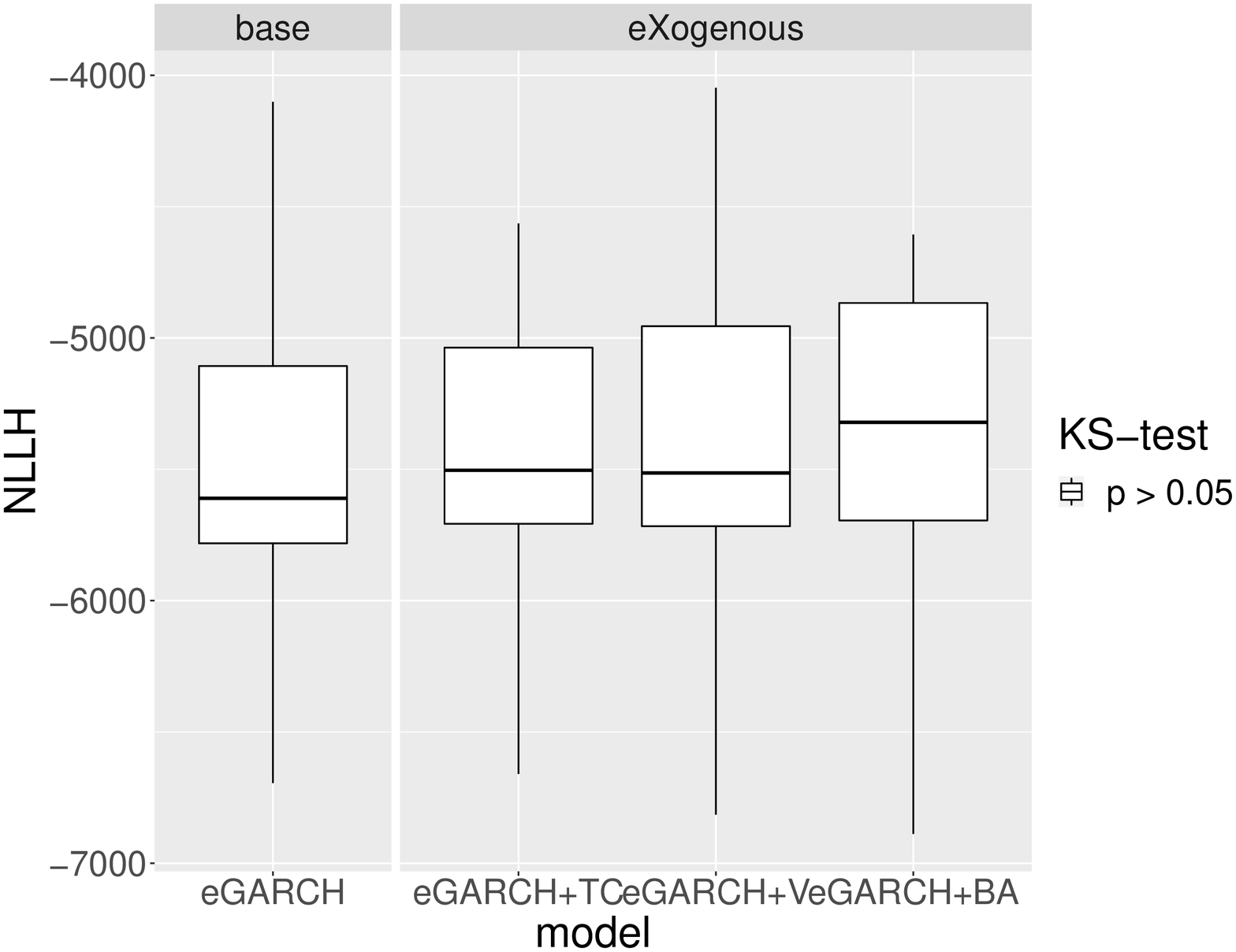}  
\end{subfigure}
\begin{subfigure}{.5\textwidth}
  \centering
  \includegraphics[width=1\linewidth]{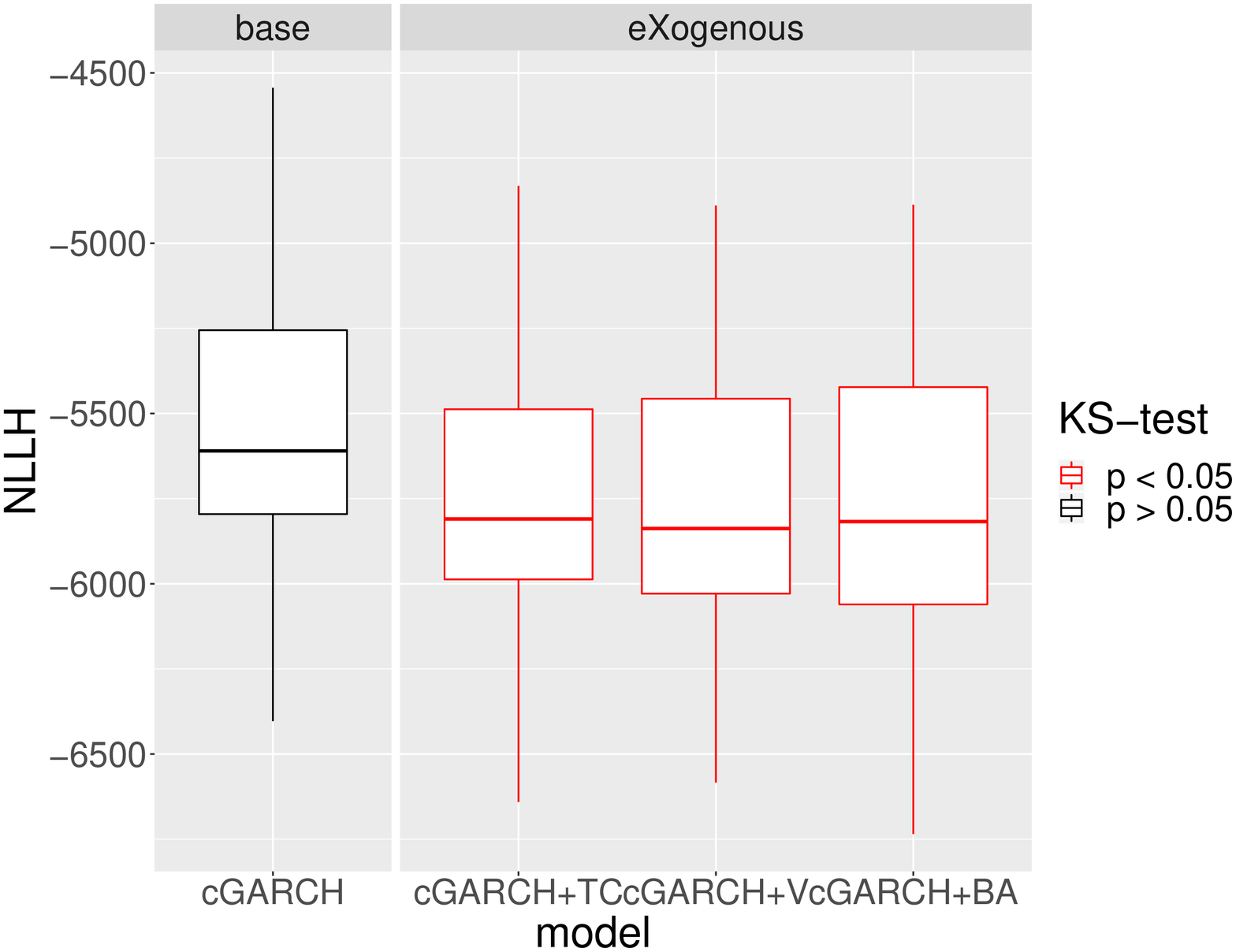}  
\end{subfigure}
\begin{subfigure}{.5\textwidth}
  \centering
  \includegraphics[width=1\linewidth]{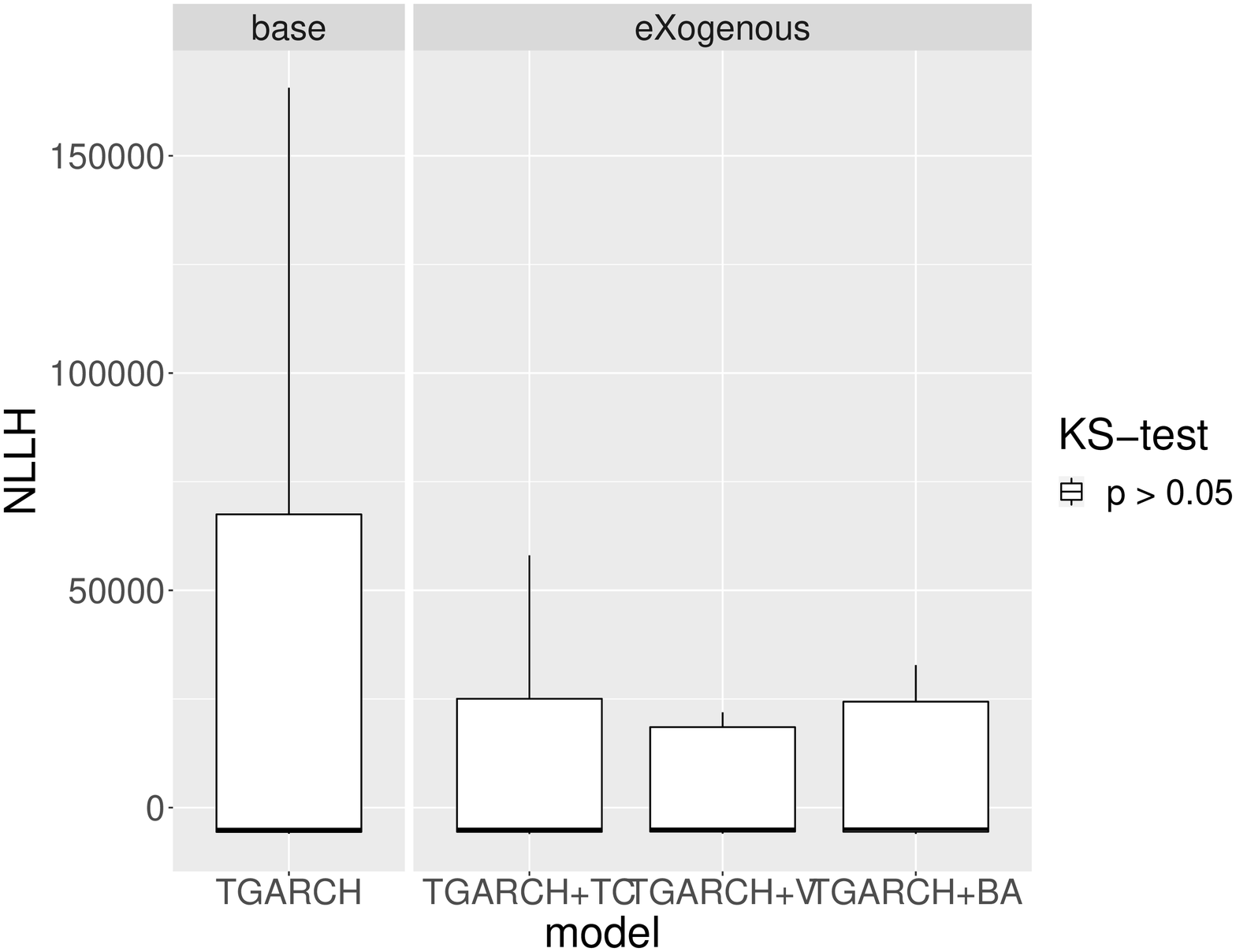}  
\end{subfigure}
\caption{Bootstrap robustness check over N = 100 splitting points with T = 1000 training points and T = 1000 test size for GARCH and GARCHX models. The price is defined as mid-quote. The non-parametric Kolmogorov–Smirnov test of the equality of the NLLH out-of-sample distributions between the GARCH and GARCHX models is done. \textbf{a)} KS test implies a significant difference for both external signals for the GARCH model. \textbf{b)} KS test implies no significant difference for external signals for eGARCH model. \textbf{c)} KS test implies no significant difference for both external signals for the cGARCH model. \textbf{d)} KS test implies no significant difference for external signals for the TGARCH model.}
\label{fig:mqnllh}
\end{figure}\\
In the Eq. 11 $M_i$ represents a model from GARCH family $\{$GARCH, cGARCH, eGARCH, TGARCH$\}$ and $M_{i,j}$ denotes its corresponding GARCHX extension, where external signal $j \in \{$Volume, Twitter, Bid-ask spread$\}$.
Models $M_i$ and $M_{i,j}$ will have empirical distribution functions $\psi_{M_i}(\Tilde{\mathcal{L}})$ and $\psi_{M_{i,j}}(\Tilde{\mathcal{L}})$, respectively (see boxplots estimates in Fig. \ref{fig:vwapnllh}). We calculate the Kolmogorov-Smirnov (KS) statistics between corresponding empirical predictive out-of-sample NLLH distributions:
\begin{equation}
KS_{i,j} = \sup_{\Tilde{\mathcal{L}}} |\psi_{M_i}(\Tilde{\mathcal{L}}) - \psi_{M_{i,j}}(\Tilde{\mathcal{L}})|,    
\end{equation}
and obtain its statistical significance. 
In Fig. \ref{fig:vwapnllh} and \ref{fig:mqnllh} we can see that both GARCH and cGARCH models show significant improvements with all the external variables and both price definitions, under the bootstrapping KS-NLLH robustness check. That is not surprising, as the non-parametric KS test is not very powerful \cite{Marozzi2013}. However, significant differences for the GARCH and cGARCH models allow us to confirm that its predictive power is robust under temporal bootstrapping conditions. 
Finally, we take the GARCH volatility process as representative and perform additional bootstrapping KS-NLLH robustness checks on two additional segments (March -- April 2019 and November -- December 2019) and we see similar results (See Fig. \ref{fig:other}, Appendix).

\section{Discussion}
Although the theoretical foundations of the effects of information on markets have been proposed a long time ago \cite{bachelier2011louis,Mandelbrot}, they were further developed in 1970, as "weak", "semi-strong", and "strong" forms of efficient market hypothesis \cite{Fama1970}.  The mathematical models of information effects continued to advance in the 70s as well, by the proposition of the Mixture of Distribution Hypothesis \cite{Clark1973}, which states that the dynamics of price returns are governed by the information flow available to the traders. Following the growth of computerized systems and the availability of empirical data in the 80s, more elaborate statistical models were proposed, such as generalized autoregressive conditional heteroscedasticity models (GARCH) \cite{GARCH} and news Poisson-jump processes \cite{jorion1988jump} with constant intensity. Furthermore, studies from the 2000s generalized the news Poisson-jump processes by introducing time-varying jump effects, supporting it with the statistical evidence of time variation in the jump size distribution \cite{chan2002conditional,maheu2004news}. \\
In this paper, we have analyzed the effects of information flow on the cryptocurrency Bitcoin exchange market that appeared with the introduction of blockchain technology in 2008 \cite{Nakamoto2008}.
Although the trading volume in the largest cryptocurrency markets has grown exponentially in the last 10 years, still the research on their (in)efficiency quantification is ongoing \cite{Tran2019,Kristoufek2019}.
We have focused on the Bitcoin, the largest cryptocurrency w.r.t. market capitalization, and used the reliable data of price returns and traded volume and bid-ask spread from Bitfinex exchange market \cite{hougan2019SEC} on a minute-level granularity.
The price returns were calculated using two different definitions, VWAP and mid-quote, to account for possible market-microstructure noise. Another reason, why we have concentrated on the Bitcoin, was the availability of Twitter-related data \cite{beck2019sensing}.
We have used the social media signals from Twitter, trading volume and bid-ask spread from the Bitcoin market as a proxy for information flow together with the GARCH family of \cite{engle2002new} processes to quantify the prediction power for the price volatility. \\
We started the analysis by employing recently developed  non-parametric information-theoretic transfer entropy measures \cite{Schreiber_2000,ETE,TEpval}, to confirm the nonlinear relationship between the exogenous proxy for information (trading volume, bid-ask spread and cryptocurrency related tweets) and squared price returns (proxy for volatility). 
Further on, we have made extensive experiments on the following models: GARCH, eGARCH, cGARCH, and TGARCH on the minute level data of price returns, Twitter volume, exchange volume data and bid-ask spread. 
Our testing procedure consisted of multi-stage statistical checks: (i) out-of-sample $R^2$ and Pearson correlation measurements, (ii) out-of-sample predictive likelihood measurements with the likelihood ratio test on 8000 points, (iii) bootstrapped predictive likelihood measurements with the non-parametric Kolmogorov-Smirnov test.
From the predictive perspective of the non-linear parametric GARCH model, we have found that exogenous proxy for information flow significantly improves out-of-sample minute volatility predictions for the GARCH and cGARCH \cite{ComponentGARCH}  models. It is not surprising that the basic GARCH model is outperforming more advanced models \cite{jafari2007does,andersen1998answering} such as eGARCH \cite{nelson1991eGARCH} and TGARCH \cite{TARCH} on out-of-sample data. 
Also, a previous study \cite{Katsiampa2017}, found that the cGARCH model on the Bitcoin market was performing the best on in-sample daily returns. \\
Finally, we have taken the GARCH model and applied the bootstrapping on two additional segments (March -- April 2019 with 38 000 points and November -- December 2019 with 52 000 points) and we observe that our observations still hold (see Appendix, Fig. \ref{fig:other}). 

\newpage

\section{Appendix}
\FloatBarrier

\begin{figure}
\centering
  \includegraphics[width=0.8\textwidth]{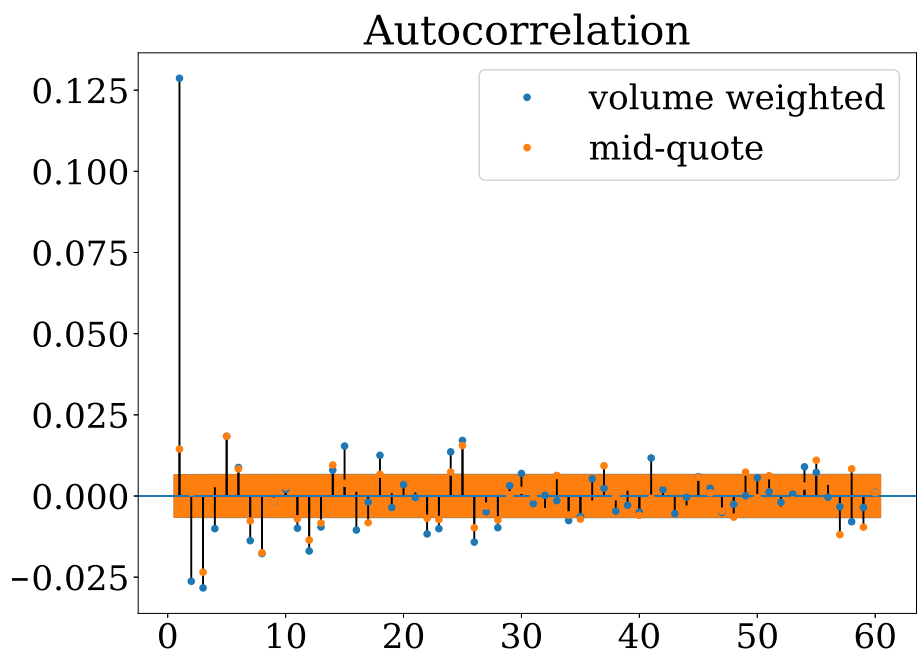}
\caption{Autocorrelation of price returns. The first order autocorrelation of mid-quote price returns is significantly smaller than of volume weighted price returns, indicated a smaller level of microstructure noise in mid-quote price returns. Confidence interval }
\label{fig:correlation}       
\end{figure}


\begin{figure}[!ht]
\centering 
\begin{subfigure}{0.5\textwidth}
  \centering
  \includegraphics[width=1\linewidth]{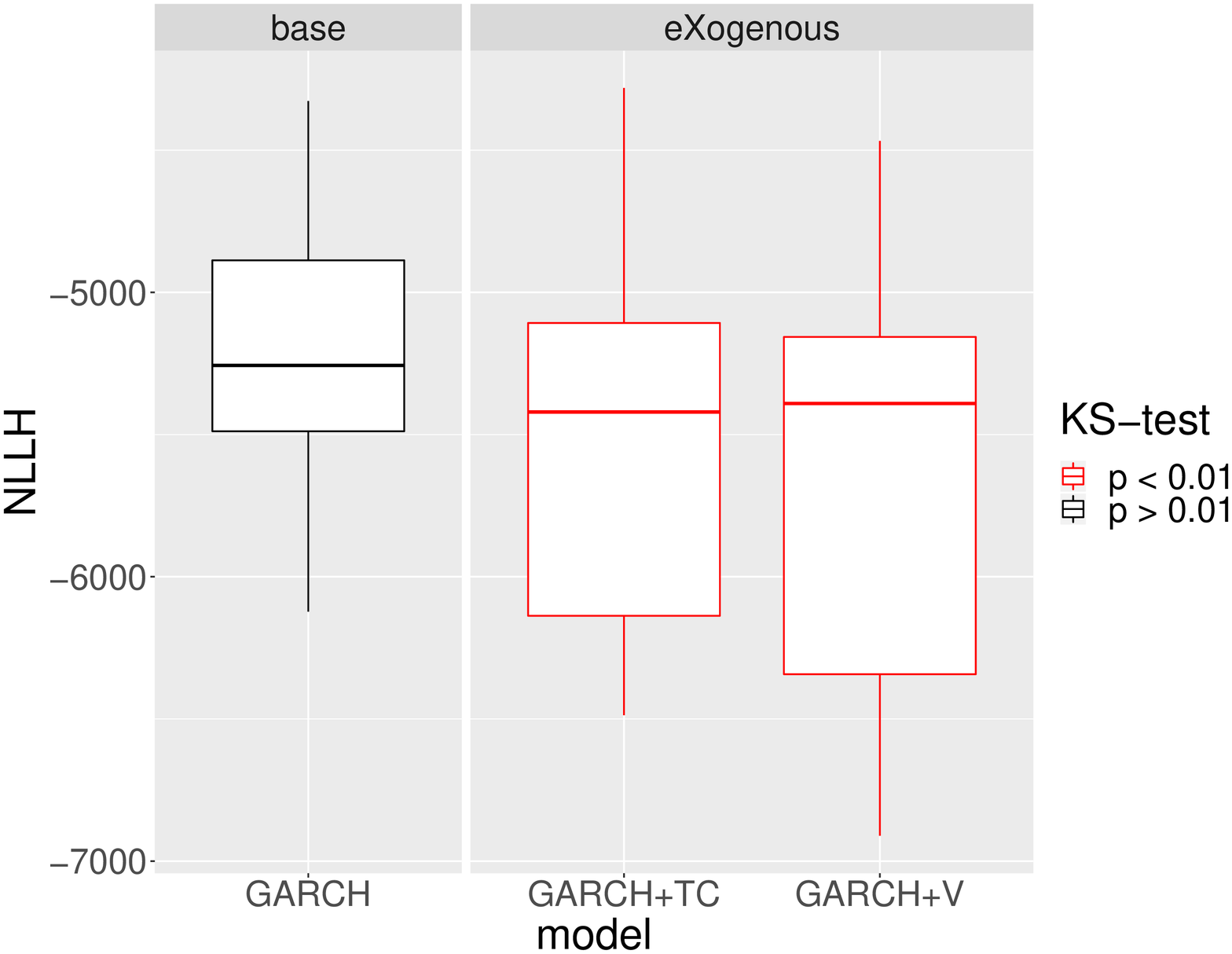}
  \caption{November 3rd 2019 to December 9th 2019}
\end{subfigure}%
\begin{subfigure}{0.5\textwidth}
  \centering
  \includegraphics[width=1\linewidth]{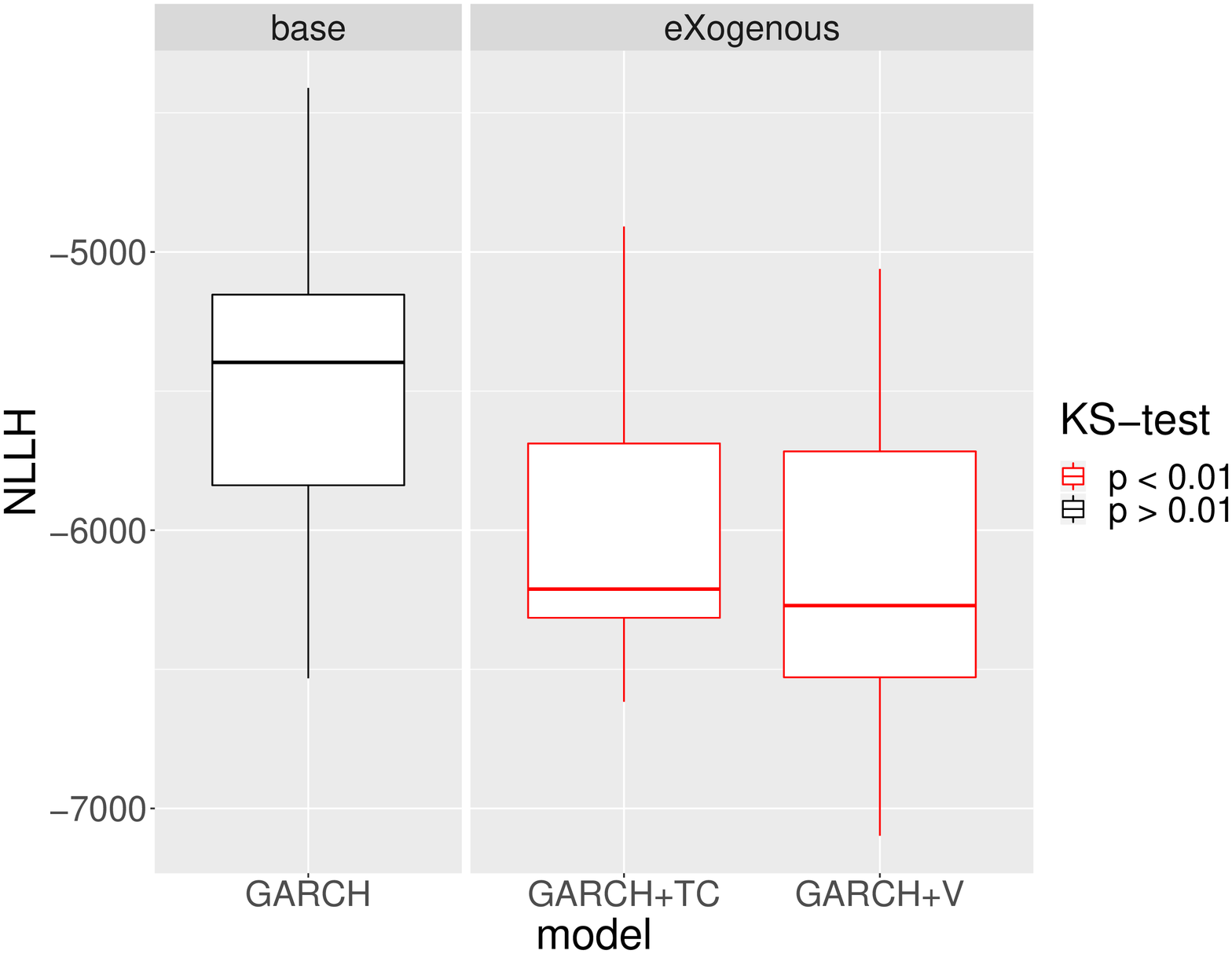}
  \caption{ March 18th 2019 to April 9th 201}
\end{subfigure}
\caption{ Bootstrap robustness check over N = 100 splitting points with T = 1000 training points and T = 1000 points in test size for GARCH and GARCHX models. The non-parametric Kolmogorov–Smirnov test of the equality of the NLLH out-of-sample distributions between GARCH and GARCHX models is done.
\textbf{a)} KS test implies a significant difference for all external signals for the GARCH model in the period from November 3rd 2019 to December 9th 2019 with 52 000 observations.
\textbf{b)} KS test implies a significant difference for all external signals for the GARCH model in the period from March 18th 2019 to April 9th 2019 with 38 000 observations.}
  \label{fig:other}

\end{figure}

\FloatBarrier
\bibliographystyle{IEEEtran}       
\bibliography{reference}   

\begin{thebibliography}{10}
\providecommand{\url}[1]{#1}
\csname url@samestyle\endcsname
\providecommand{\newblock}{\relax}
\providecommand{\bibinfo}[2]{#2}
\providecommand{\BIBentrySTDinterwordspacing}{\spaceskip=0pt\relax}
\providecommand{\BIBentryALTinterwordstretchfactor}{4}
\providecommand{\BIBentryALTinterwordspacing}{\spaceskip=\fontdimen2\font plus
\BIBentryALTinterwordstretchfactor\fontdimen3\font minus
  \fontdimen4\font\relax}
\providecommand{\BIBforeignlanguage}[2]{{%
\expandafter\ifx\csname l@#1\endcsname\relax
\typeout{** WARNING: IEEEtran.bst: No hyphenation pattern has been}%
\typeout{** loaded for the language `#1'. Using the pattern for}%
\typeout{** the default language instead.}%
\else
\language=\csname l@#1\endcsname
\fi
#2}}
\providecommand{\BIBdecl}{\relax}
\BIBdecl

\bibitem{bachelier2011louis}
L.~Bachelier, \emph{Louis Bachelier's theory of speculation: the origins of
  modern finance}.\hskip 1em plus 0.5em minus 0.4em\relax Princeton University
  Press, 2011.

\bibitem{Mandelbrot}
B.~Mandelbrot, ``The variation of certain speculative prices,'' \emph{The
  Journal of Business}, vol.~36, no.~4, pp. 394--419, 1963.

\bibitem{stanley2000introduction}
H.~Stanley and R.~Mantegna, \emph{An introduction to econophysics}.\hskip 1em
  plus 0.5em minus 0.4em\relax Cambridge University Press, Cambridge, 2000.

\bibitem{Clark1973}
P.~K. Clark, ``A subordinated stochastic process model with finite variance for
  speculative prices,'' \emph{Econometrica}, vol.~41, no.~1, p. 135, Jan. 1973.

\bibitem{ARCH}
R.~F. Engle, ``Autoregressive conditional heteroscedasticity with estimates of
  the variance of united kingdom inflation,'' \emph{Econometrica: Journal of
  the Econometric Society}, pp. 987--1007, 1982.

\bibitem{GARCH}
T.~Bollerslev, ``Generalized autoregressive conditional heteroskedasticity,''
  \emph{Journal of econometrics}, vol.~31, no.~3, pp. 307--327, 1986.

\bibitem{jorion1988jump}
P.~Jorion, ``On jump processes in the foreign exchange and stock markets,''
  \emph{The Review of Financial Studies}, vol.~1, no.~4, pp. 427--445, 1988.

\bibitem{chan2002conditional}
W.~H. Chan and J.~M. Maheu, ``Conditional jump dynamics in stock market
  returns,'' \emph{Journal of Business \& Economic Statistics}, vol.~20, no.~3,
  pp. 377--389, 2002.

\bibitem{maheu2004news}
J.~M. Maheu and T.~H. McCurdy, ``News arrival, jump dynamics, and volatility
  components for individual stock returns,'' \emph{The Journal of Finance},
  vol.~59, no.~2, pp. 755--793, 2004.

\bibitem{BTCbook}
D.~Chuen, \emph{Handbook of Digital Currency: Bitcoin, Innovation, Financial
  Instruments, and Big Data}.\hskip 1em plus 0.5em minus 0.4em\relax Academic
  Press, 2015.

\bibitem{Nakamoto2008}
S.~Nakamoto, ``Bitcoin: A peer-to-peer electronic cash system,'' 2008.

\bibitem{gandal2017price}
N.~Gandal, J.~Hamrick, T.~Moore, and T.~Oberman, ``Price manipulation in the
  bitcoin ecosystem,'' 2017.

\bibitem{ciaian2016economics}
P.~Ciaian, M.~Rajcaniova, and d.~Kancs, ``The economics of bitcoin price
  formation,'' \emph{Applied Economics}, vol.~48, pp. 1799--1815, 2016.

\bibitem{cheah2015speculative}
E.-T. Cheah and J.~Fry, ``Speculative bubbles in bitcoin markets? an empirical
  investigation into the fundamental value of bitcoin,'' \emph{Economics
  Letters}, vol. 130, pp. 32--36, 2015.

\bibitem{kristoufek2015main}
L.~a. Kristoufek, ``What are the main drivers of the bitcoin price? evidence
  from wavelet coherence analysis,'' \emph{PloS one}, vol.~10, no.~4, p.
  e0123923, 2015.

\bibitem{BouchaudBTC}
J.~Donier and J.-P. Bouchaud, ``Why do markets crash? bitcoin data offers
  unprecedented insights,'' \emph{PLOS ONE}, vol.~10, pp. 1--11, 2015.

\bibitem{sornetteBTC}
S.~Wheatley, D.~Sornette, T.~Huber, M.~Reppen, and R.~N. Gantner, ``Are bitcoin
  bubbles predictable? combining a generalized metcalfe's law and the lppls
  model,'' 2018.

\bibitem{Katsiampa2017}
P.~Katsiampa, ``Volatility estimation for bitcoin: A comparison of {GARCH}
  models,'' \emph{Economics Letters}, vol. 158, pp. 3--6, 2017.

\bibitem{GuoBifetAntulov18}
T.~{Guo}, A.~{Bifet}, and N.~{Antulov-Fantulin}, ``Bitcoin volatility
  forecasting with a glimpse into buy and sell orders,'' in \emph{2018 IEEE
  International Conference on Data Mining (ICDM)}, Nov 2018, pp. 989--994.

\bibitem{Ron2013BTC}
D.~Ron and A.~Shamir, ``Quantitative analysis of the full bitcoin transaction
  graph,'' in \emph{International Conference on Financial Cryptography and Data
  Security}.\hskip 1em plus 0.5em minus 0.4em\relax Springer, 2013, pp. 6--24.

\bibitem{ElBahrawy2017}
A.~ElBahrawy, L.~Alessandretti, A.~Kandler, R.~Pastor-Satorras, and
  A.~Baronchelli, ``Evolutionary dynamics of the cryptocurrency market,''
  \emph{Royal Society Open Science}, vol.~4, no.~11, p. 170623, 2017.

\bibitem{AntTol18}
N.~Antulov-Fantulin, D.~Tolic, M.~Piskorec, Z.~Ce, and I.~Vodenska, ``Inferring
  short-term volatility indicators from the bitcoin blockchain,'' in
  \emph{Complex Networks and Their Applications VII}.\hskip 1em plus 0.5em
  minus 0.4em\relax Cham: Springer International Publishing, 2019, pp.
  508--520.

\bibitem{Hayes2015}
A.~Hayes, ``Cryptocurrency value formation: An empirical analysis leading to a
  cost of production model for valuing bitcoin.'' \emph{{SSRN} Electronic
  Journal}, 2015.

\bibitem{Bolt2016}
W.~Bolt, ``On the value of virtual currencies,'' \emph{{SSRN} Electronic
  Journal}, 2016.

\bibitem{nadarajah2017inefficiency}
S.~Nadarajah and J.~Chu, ``On the inefficiency of bitcoin,'' \emph{Economics
  Letters}, vol. 150, pp. 6--9, 2017.

\bibitem{kristoufek2013bitcoin}
L.~Kristoufek, ``Bitcoin meets google trends and wikipedia: Quantifying the
  relationship between phenomena of the internet era,'' \emph{Scientific
  reports}, vol.~3, p. 3415, 2013.

\bibitem{li2018sentiment}
T.~R. Li, A.~S. Chamrajnagar, X.~R. Fong, N.~R. Rizik, and F.~Fu,
  ``Sentiment-based prediction of alternative cryptocurrency price fluctuations
  using gradient boosting tree model,'' \emph{arXiv preprint arXiv:1805.00558},
  2018.

\bibitem{kim2016predicting}
Y.~B. Kim, J.~G. Kim, W.~Kim, J.~H. Im, T.~H. Kim, S.~J. Kang, and C.~H. Kim,
  ``Predicting fluctuations in cryptocurrency transactions based on user
  comments and replies,'' \emph{PloS one}, vol.~11, 2016.

\bibitem{Garcia2015}
D.~Garcia and F.~Schweitzer, ``Social signals and algorithmic trading of
  bitcoin,'' \emph{Royal Society Open Science}, vol.~2, no.~9, p. 150288, 2015.

\bibitem{zero}
\BIBentryALTinterwordspacing
J.~Gatheral and R.~Oomen, ``{Zero-intelligence realized variance estimation},''
  \emph{Finance and Stochastics}, vol.~14, no.~2, pp. 249--283, April 2010.
  [Online]. Available:
  \url{https://ideas.repec.org/a/spr/finsto/v14y2010i2p249-283.html}
\BIBentrySTDinterwordspacing

\bibitem{beck2019sensing}
J.~Beck, R.~Huang, D.~Lindner, T.~Guo, Z.~Ce, D.~Helbing, and
  N.~Antulov-Fantulin, ``Sensing social media signals for cryptocurrency
  news,'' in \emph{Companion Proceedings of The 2019 World Wide Web
  Conference}, 2019, pp. 1051--1054.

\bibitem{Tauchen}
G.~E. Tauchen and M.~Pitts, ``The price variability-volume relationship on
  speculative markets,'' \emph{Econometrica}, vol.~51, no.~2, pp. 485--505,
  1983.

\bibitem{Schreiber_2000}
T.~Schreiber, ``Measuring information transfer,'' \emph{Physical Review
  Letters}, vol.~85, no.~2, p. 461–464, Jul 2000.

\bibitem{Barnett_2009}
\BIBentryALTinterwordspacing
L.~Barnett, A.~B. Barrett, and A.~K. Seth, ``Granger causality and transfer
  entropy are equivalent for gaussian variables,'' \emph{Physical Review
  Letters}, vol. 103, no.~23, Dec 2009. [Online]. Available:
  \url{http://dx.doi.org/10.1103/PhysRevLett.103.238701}
\BIBentrySTDinterwordspacing

\bibitem{TEpval}
T.~Dimpfl and F.~J. Peter, ``Using transfer entropy to measure information
  flows between financial markets,'' \emph{Studies in Nonlinear Dynamics and
  Econometrics}, vol.~17, no.~1, pp. 85--102, 2013.

\bibitem{ETE}
R.~Marschinski and H.~Kantz, ``Analysing the information flow between financial
  time series,'' \emph{The European Physical Journal B-Condensed Matter and
  Complex Systems}, vol.~30, no.~2, pp. 275--281, 2002.

\bibitem{nelson1991eGARCH}
D.~Nelson, ``Conditional heteroskedasticity in asset returns: A new approach,''
  \emph{Econometrica}, vol.~59, no.~2, pp. 347--70, 1991.

\bibitem{TARCH}
J.-M. Zakoian, ``Threshold heteroskedastic models,'' \emph{Journal of Economic
  Dynamics and control}, vol.~18, no.~5, pp. 931--955, 1994.

\bibitem{1999cGARCH}
G.~Lee and R.~Engle, ``A permanent and transitory component model of stock
  return volatility,'' \emph{Cointegration, Causality and Forecasting: A
  Festschrift in Honor of Clive W.J. Granger}, pp. 475--497, 1999.

\bibitem{andersen1998answering}
T.~G. Andersen and T.~Bollerslev, ``Answering the skeptics: Yes, standard
  volatility models do provide accurate forecasts,'' \emph{International
  economic review}, pp. 885--905, 1998.

\bibitem{wu2014gaussian}
Y.~Wu, J.~M. Hern{\'a}ndez-Lobato, and Z.~Ghahramani, ``Gaussian process
  volatility model,'' in \emph{NIPS}, 2014, pp. 1044--1052.

\bibitem{Marozzi2013}
M.~Marozzi, ``Nonparametric simultaneous tests for location and scale testing:
  A comparison of several methods,'' \emph{Communications in Statistics -
  Simulation and Computation}, vol.~42, no.~6, pp. 1298--1317, Jul. 2013.

\bibitem{Fama1970}
E.~F. Fama, ``Efficient capital markets: A review of theory and empirical
  work,'' \emph{The Journal of Finance}, vol.~25, no.~2, p. 383, May 1970.

\bibitem{Tran2019}
V.~L. Tran and T.~Leirvik, ``Efficiency in the markets of crypto-currencies,''
  \emph{Finance Research Letters}, p. 101382, Nov. 2019.

\bibitem{Kristoufek2019}
L.~Kristoufek and M.~Vosvrda, ``Cryptocurrencies market efficiency ranking: Not
  so straightforward,'' \emph{Physica A: Statistical Mechanics and its
  Applications}, vol. 531, p. 120853, Oct. 2019.

\bibitem{hougan2019SEC}
M.~Hougan, H.~Kim, M.~Lerner, and B.~A. Management, ``Economic and non-economic
  trading in bitcoin: Exploring the real spot market for the world’s first
  digital commodity,'' \emph{Bitwise Asset Management}, 2019.

\bibitem{engle2002new}
R.~Engle, ``New frontiers for arch models,'' \emph{Journal of Applied
  Econometrics}, vol.~17, no.~5, pp. 425--446, 2002.

\bibitem{ComponentGARCH}
R.~F. Engle and M.~E. Sokalska, ``Forecasting intraday volatility in the us
  equity market. multiplicative component garch,'' \emph{Journal of Financial
  Econometrics}, vol.~10, no.~1, pp. 54--83, 2012.

\bibitem{jafari2007does}
G.~Jafari, A.~Bahraminasab, and P.~Norouzzadeh, ``Why does the standard garch
  (1, 1) model work well?'' \emph{International Journal of modern physics C},
  vol.~18, no.~07, pp. 1223--1230, 2007.

\end{thebibliography}

\end{document}